\documentclass[twosided,11pt]{article}

\usepackage[round]{natbib}
\usepackage{color}
\usepackage{graphics}
\usepackage{graphicx}
\usepackage{physics}
\usepackage{tikz}
\usepackage{xcolor}
\usepackage{subcaption}
\usepackage{tabularx}
\usepackage{booktabs}
\usepackage{epigraph}

\definecolor{lightblue}{RGB}{102, 178, 255}
\definecolor{darkblue}{RGB}{51, 102, 204}
\definecolor{lightgreen}{RGB}{102, 204, 102}
\definecolor{darkred}{RGB}{204, 51, 51}

\usepackage{amsmath}

\usepackage{amsmath,amssymb,amsfonts,amsthm}

\newtheorem{conjecture}{Conjecture}
\newtheorem{remark}{Remark}

\usepackage{mathtools} 

\usepackage[noend]{algpseudocode}
 \usepackage[linesnumbered, ruled, vlined]{algorithm2e}

\usepackage{multirow}

\usepackage{caption}

\begin{document}

\title{
Bayes or Heisenberg:   Who(se) Rules?
}

\author{Volker Tresp\footnote{Corresponding Author:  Volker.Tresp@lmu.de.}, Hang Li, Federico Harjes, and Yunpu Ma \\ 
LMU Munich}


 \maketitle

\begin{abstract}

Although quantum systems are generally described by quantum state vectors, we show that in certain cases their measurement processes can be reformulated as probabilistic equations expressed in terms of probabilistic state vectors. These probabilistic representations can, in turn, be approximated by the neural network dynamics of the Tensor Brain (TB) model.

The Tensor Brain is a recently proposed framework for modeling perception and memory in the brain, providing a biologically inspired mechanism for efficiently integrating generated symbolic representations into reasoning processes.  

We begin by reviewing projective and generalized measurements in quantum systems. In the probabilistic quantum formulation, a quantum state is replaced by a probabilistic state, and unitary transformations are represented by unitary-stochastic matrices. Building on this correspondence, probabilistic quantum algorithms can be implemented using \textit{pro-bits}---the probabilistic analogue of qubits---together with networks of unitary-stochastic gates. By introducing a series of controlled approximations, we derive a neural network formulation that constitutes the foundation of the Tensor Brain algorithms.

Quantum systems exhibit a distinctive generative measurement process: measuring a quantum state not only produces an outcome but also alters the state itself. In the Tensor Brain framework, analogous measurements are performed on the cognitive brain state (CBS). The possible outcomes—such as concepts or past time instances—feed back into and modify the CBS. 

We further examine the relationship between the Tensor Brain and the Bayesian Brain, identifying a special case where probabilistic quantum reasoning reduces exactly to Bayesian reasoning. In the general case, however, we argue that the direct application of Bayes’ rule often becomes computationally intractable, while probabilistic quantum algorithms remain tractable.

Finally, we connect the Tensor Brain framework to transformer-based large language models (LLMs), emphasizing both their similarities and differences. In this view, memory operations in the Tensor Brain parallel Retrieval-Augmented Generation (RAG) in LLMs. Moreover, skip connections can be interpreted as Bayesian priors, while attention mechanisms correspond to ignorant probabilistic quantum measurements.

\end{abstract}
 
\tableofcontents

\mbox{}

\mbox{}


\epigraph{In classical physics, I don't know where my marbles are, but Mother Nature does; in quantum physics: we both lost our marbles.}{VT}

\section{Introduction}

How does information processing in the brain work? How do perception and memory emerge? And should we consider fundamentally new approaches to modeling and reasoning in AI? The Tensor Brain (TB) \citep{tresp2023tensor,tresp2025tensor} is an attempt to provide such answers. TB is formulated as a nonlinear state-space model, in which the system state is represented by the activation of formal neurons in the brain’s representation layer. We refer to this activation vector as the cognitive brain state (CBS), which is closely related to the concept of the global workspace in consciousness research \citep{baars1997theater,dehaene2014consciousness}.



While the Cognitive Brain State (CBS) is distributed and subsymbolic, neurons in a second layer—the index layer—encode symbolic interpretations. Perception and memory emerge from the interaction between these two layers.
The resulting processing behavior is non-standard, particularly in its treatment of top-down inference and the generation of symbolic labels.

During an integration phase, multiple indices may be activated simultaneously, reflecting a convergence of cognitive representations. This is followed by an evolution phase, in which a third layer, i.e., the dynamic context layer,  compresses the current CBS and governs the transition to the next state, thereby shaping the system’s temporal dynamics.

This paper builds on the TB framework and investigates quantum theory as a potential foundation for information processing in the brain. From this perspective, the CBS corresponds to a quantum state expressed in the computational basis and implemented with qubits, while the index layer corresponds to the measurement basis.

We introduce the Heisenberg–Bayes Positive Operator-Valued Measure (HB-POVM), a measurement scheme that preserves state information, and examine the role of postselection, which enables the system to restrict itself to a smaller set of known symbols.

We argue that measurement outcomes in quantum systems are inherently probabilistic and propose a specific measurement protocol reflecting this interpretation. Within this framework, the quantum state can be replaced by a probabilistic state, and both evolution and measurement operations can be expressed probabilistically through unitary-stochastic matrices, which are a special case of doubly stochastic matrices.

A key result is that with unitary-stochastic operator matrices, a probabilistic HB-POVM update is mathematically equivalent to a Bayesian update in a generative hidden Markov model (gHMM). Once postselection is applied, however, this equivalence no longer holds: probabilistic HB-POVM inference remains tractable, whereas gHMM inference becomes intractable.

We further show that pro-bits—probabilistic bits obtained by tensorizing the probabilistic state space—can be implemented as neurons within the TB representation layer. By introducing a series of approximations, we derive scalable TB algorithms whose evidence updates are computationally more efficient than standard Bayesian updates.

The paper is organized as follows.
In the next section, we review relevant literature on quantum probability theory, quantum decoherence, the quantum-brain hypothesis, quantum cognition, the Bayesian brain hypothesis, and the TB framework.

Section~\ref{sec:tbt} presents the core equations of the TB. The TB is formulated as a nonlinear state-space model in which the system state is represented by the activation of formal neurons in the representation layer. We refer to this activation vector as the cognitive brain state (CBS), which will later be related to both the quantum state and the probabilistic state. Whereas the CBS is distributed and subsymbolic, neurons in the index layer carry a symbolic interpretation.

From the TB, we propose three general principles:
\begin{itemize}
\item Top-down inference is essential for both memory functions and symbol grounding.
\item Information processing in the brain emerges from the interaction of two layers: the subsymbolic representation layer and the symbolic index layer.
\item The TB—and possibly the brain itself—operates as a sampling engine.
\end{itemize}

Section~\ref{sec:qtheory} provides background on quantum theory. We describe the computational basis ($X$-basis), associated with the CBS, and the $Y$-basis, associated with symbolic indices. We also review PVMs, POVMs, and postselection, and introduce the Heisenberg–Bayes POVM (HB-POVM). Two main results are emphasized:
\begin{itemize}
\item The HB-POVM is a novel contribution, linking quantum measurements with Bayesian measurements.
\item A qubit representation arises naturally through tensorization of the state space.
\end{itemize}

Section~\ref{sec:probq} introduces probabilistic quantum models. We present a probabilistic measurement protocol in which quantum states are replaced by probabilistic states and unitary matrices by unitary-stochastic matrices. The main conclusions are:
\begin{itemize}
\item A probabilistic HB-POVM is equivalent to inference in a generative hidden Markov model (gHMM) in the absence of postselection.
\item With postselection, HB-POVM inference remains tractable, in contrast to gHMM inference.
\item Pro-bits and unitary-stochastic gates enable probabilistic computation through stochastic sampling.
\end{itemize}

Section~\ref{sec:fact} presents a neural network implementation. We approximate pro-bits using formal neurons, applying independence assumptions and Jensen’s approximation. Key results include:
\begin{itemize}
\item The neural HB-POVM update is equivalent to a neural PVM update plus a skip connection, providing an interpretation of skip connections  as logit priors.
\item Index embedding vectors constitute the connection weights linking the embedding layer to the index layer.
\end{itemize}

In Section~\ref{sec:back}, we discuss TB operations for perception, semantic memory, and episodic memory:

\begin{itemize}
\item The CBS corresponds to firing rates in the representation layer.
\item Memory integration in the TB parallels Retrieval-Augmented Generation (RAG) in large language models (LLMs).
\item We propose a new interpretation of attention: as a measurement with unknown outcomes in the $Y$-basis.
\item We propose that the   dynamic context layer might be the basis for a working memory. 
\end{itemize}

Section \ref{sec:discuss} examine the relationship of the TB to transformer-based LLMs, explicit versus distributed indices, and quantum cognition.
Section~\ref{sec:concl} presents  final conclusions.


\section{Related Work}

\subsection{Quantum Theory}

The foundations of quantum mechanics were established by pioneers such as Werner Heisenberg, Erwin Schrödinger, Max Born, Pascual Jordan, Wolfgang Pauli, and Paul Dirac. \cite{von2018mathematical} provided the first rigorous mathematical formulation, introducing Hilbert spaces, operators, and density matrices, which remain the cornerstone of quantum probability theory.

Quantum information theory emphasizes the computational aspects of quantum mechanics and serves as the basis for quantum computing. Although its primary focus is not on the study of natural phenomena, quantum computing addresses the engineering of quantum circuits to solve computational problems. In addition to standard projective measurements (PVMs), quantum computing employs generalized measurements (POVMs)—both of which play a central role in this paper. For further background, we frequently refer to the textbook by \cite{nielsen2010quantum}.

In this work, we adopt the Copenhagen interpretation of quantum theory. Alternative interpretations also exist, including the many-worlds interpretation, the de Broglie–Bohm interpretation, and quantum Bayesianism (QBism), the latter encompassing a family of Bayesian-inspired perspectives on quantum mechanics.

Our approach builds directly on the mathematical foundations of quantum theory: we reinterpret projective and generalized measurements within a probabilistic framework, thereby establishing a connection between quantum and probabilistic reasoning.

 \subsection{Quantum Decoherence}
 \label{sec:qdec}

Quantum decoherence carries two distinct meanings \cite[p.~398]{nielsen2010quantum}. First, in the context of quantum computation and quantum information, it refers to any noise process affecting quantum operations. Modeling the influence of the environment on a quantum system naturally leads to generalized measurements and the POVM framework.

Second, certain interpretations suggest that the measurement postulate 
(i.e., postulate~3 of quantum mechanics described in Section~\ref{sec:pvm}) becomes redundant if the measuring apparatus and environment are explicitly included in the model \citep{zeh1970interpretation}. Decoherence arises when a quantum system interacts with its environment, forming a high-dimensional joint state. The subsystem of interest then appears to follow classical probability rules rather than quantum mechanics, as coherence (phase information) is effectively destroyed.

Technically, decoherence suppresses the off-diagonal elements of the reduced density matrix \citep{tegmark2000importance}. Once the density matrix becomes diagonal, the system behaves probabilistically. The concept was first introduced by \citep{zeh1970interpretation}, extending Bohm’s earlier ideas \citep{bohm1952suggested}, and later developed further through density-matrix dynamics by \citep{zurek1981pointer}.

The system studied in this paper can likewise be described probabilistically. Diagonalization of the density matrix plays a central role in our framework, particularly in the context of ignorant measurements with unknown outcomes. While we adopt the classical interpretation of quantum theory, the role of an observer in the measurements 
may be assumed by the  decoherence processes.

This analogy to decoherence motivates our replacement of quantum states with probabilistic states, which forms the foundation of the Tensor Brain algorithms.

\subsection{Quantum Computing  and the Brain}

Several researchers have proposed that the brain functions as a coherent quantum computer. \cite{penrose2011consciousness} suggested that quantum computations take place in microtubules, structural components of the neuronal cytoskeleton. According to this view, microtubules are interconnected across dendrites via gap junctions, forming a coherent, entangled quantum state spanning the cortex.

However, this theory has faced strong criticism. Analyses of decoherence times in the brain indicate that quantum coherence is lost extremely rapidly—far too quickly to support meaningful quantum computations \citep{tegmark2000importance}.

\subsection{Quantum Cognition}
\label{sec:qcog}

While classical probability theory provides a rational framework for inference, human judgment often deviates from its predictions \citep{kahneman2011thinking}. The field of quantum cognition argues that the mathematical structures of quantum theory can offer a more accurate account of human reasoning than classical models.

Quantum cognition employs a non-commutative algebra in which the ordering of events matters:
\[
P(A, \textit{and then } B) \neq P(B, \textit{and then } A),
\]
capturing phenomena such as order effects and violations of the law of total probability. Importantly, quantum cognition does not posit microscopic quantum processes in the brain; rather, it treats quantum theory as a formal modeling framework for human thought. A comprehensive overview can be found in \citep{busemeyer2012quantum}.

\subsection{The Tensor Brain and the Bayesian Brain}

The Bayesian brain hypothesis posits that the brain implements Bayesian reasoning across multiple levels of cognition \citep{dayan1995helmholtz, rao1999predictive, knill2004bayesian, kording2004bayesian, tenenbaum2006theory, griffiths2008bayesian, friston2010free}. In practice, exact Bayesian inference is computationally intractable and typically requires variational approximations. This limitation has led to criticism: Bayesian inference may be biologically implausible due to its high computational demands \citep{kwisthout2020computational, rutar2022structure}.  

The {Tensor Brain (TB)} was introduced as a biologically plausible and computationally efficient alternative \citep{tresp2023tensor, tresp2025tensor}. It has been proposed as a model for perception and memory, and it will be referenced extensively throughout this paper.  

The Tensor Brain provides the computational framework for our approach. By linking it to probabilistic quantum inference, we obtain a scalable model that unifies Bayesian, quantum, and neural perspectives.

\section{Tensor Brain (TB)}
\label{sec:tbt}

 \begin{figure}[hbt!]
\captionsetup{singlelinecheck=off}
\begin{center}
  \includegraphics[width=\linewidth]{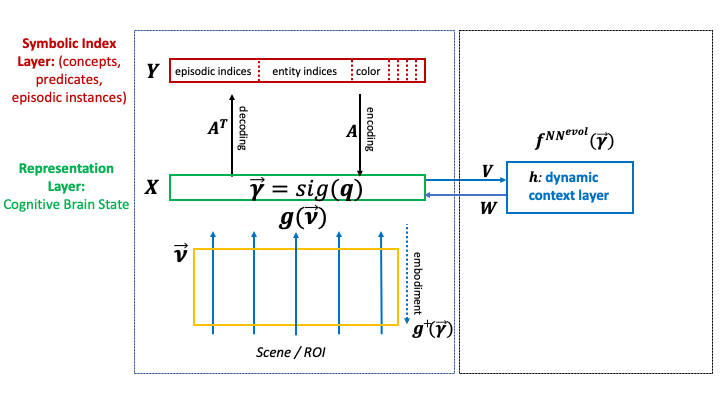}
\end{center}
\caption[foo bar]{Overall architecture. The scene input at the bottom is processed by 
several processing layers denoted by $\vec \nu$.  $\vec \nu$ denotes low-level cortical areas such as  primary sensory cortices,  early sensory regions, and maybe even subcortical regions. 
$\vec \nu$ feeds into 
the representation layer $X$, while the top layer $Y$ encodes symbolic indices.  The system’s evolution is governed by a neural network (shown on the right), with $\mathbf{h}$ denoting the hidden layer within the network. $\mathbf{g}^+(\cdot)$ symbolizes embodiment (see Section~\ref{sec:exti2}).
%
}
\label{fig:Arch}
\end{figure}

\subsection{The Algorithms}

The Tensor Brain (TB) algorithms were introduced in \citep{tresp2023tensor, tresp2025tensor}. They specify how the {cognitive brain state (CBS)} evolves, integrates external inputs, and generates symbolic outputs. The overall architecture is illustrated in Figure~\ref{fig:Arch}.  

Formally, let $\mathbf{q} \in \mathbb{R}^n$ denote the vector of pre-activations. Its post-activation vector is given by the logistic function
\[
\vec{\gamma} = \mathrm{sig}(\mathbf{q})
\] 
which we identify as the CBS. The pre-activation vector $\mathbf{q}$ is sometimes referred to as the pre-CBS. The CBS is closely related to the global workspace concept in consciousness research \citep{baars1997theater,dehaene2014consciousness}.

The TB framework is defined by three core algorithms:

\begin{itemize} 

\item \textbf{Algorithm~\ref{alg:cap-1NN}} describes the evolution of the pre-CBS. This evolution is implemented via a neural network with a hidden layer whose post-activation vector is 
$\mathbf{h} \in \mathbb{R}^{n_H}$, where $n_H$ denotes the number of hidden neurons.  

\item \textbf{Algorithm~\ref{alg:cap-2NN}} describes the integration of  $\vec \nu$, which is mapped as $\mathbf{g}(\vec \nu) \in \mathbb{R}^n$. 
Here, $\vec \nu$ denotes low-level cortical areas such as  primary sensory cortices,  early sensory regions, and maybe even subcortical regions. In the further discussion we typically assume that these are brain states driven by visual or other sensory inputs, such as V1 and V2.

\item \textbf{Algorithm~\ref{alg:cap-3NN}} assigns a label $k$ and adds the corresponding embedding vector $\mathbf{a}_k$ to the pre-CBS, where $\mathbf{a}_k$ is the $k$-th column of the matrix $\mathbf{A} \in \mathbb{R}^{n \times N_S}$, and $N_S$ denotes the number of symbolic indices. In the TB, $\alpha = \beta = 1$. 

\end{itemize}

For a given concept of interest, e.g., a particular object in a scene, Algorithm~\ref{alg:cap-1NN} is executed first, followed by Algorithm~\ref{alg:cap-2NN}.  
Finally, {Algorithm~\ref{alg:cap-3NN}} is executed, potentially multiple times, producing outcomes. Each outcome $k$ corresponds to a distinct symbolic index.  

Together, these algorithms enable the TB to process perception, integrate information, and generate symbolic representations.  


The integration phase is followed by another evolution operation (Algorithm~\ref{alg:cap-1NN}), in which the dynamic context layer (the $\mathbf{h}$-layer) summarizes the accumulated information and temporarily deactivates all symbolic indices. This allows the system to either shift attention to a new concept or maintain focus on the current one. After labeling a subject, Algorithm~\ref{alg:cap-1NN} can be invoked again to redirect focus toward another object or to infer relations between objects—such as triples of the form $(\text{subject}, \text{object}, \text{predicate})$.

\begin{algorithm}[H]
\SetAlgoLined
\DontPrintSemicolon
\caption{TB Evolution (implementing $f^{\textit{NN}^{\textit{evol}} }(\vec \gamma)$)}
\label{alg:cap-1NN}

\KwIn{pre-CBS $\mathbf{q} \in \mathbb{R}^n$; neural network parameters 
$\mathbf{V} \in \mathbb{R}^{n_H \times n}$, $\mathbf{v}_0 \in \mathbb{R}^{n_H}$, $\mathbf{W} \in \mathbb{R}^{n \times n_H}$}
\KwOut{Updated pre-CBS $\mathbf{q}$ after TB evolution}

$\mathbf{h} \gets \mathrm{sig}(\mathbf{v}_0 + \mathbf{V} \mathbf{q})$ \Comment{Hidden state (post-activation)}  

$\mathbf{q} \gets \mathbf{W} \mathbf{h}$  

\Return{$\mathbf{q}$}
\end{algorithm}

\begin{algorithm}[H]
\SetAlgoLined
\DontPrintSemicolon
\caption{Tensor Brain (TB) Input and Attention}
\label{alg:cap-2NN}

\KwIn{pre-CBS $\mathbf{q} \in \mathbb{R}^n$; embedding matrix 
$\mathbf{A} \in \mathbb{R}^{n \times N_S}$; bias vector $\mathbf{a}_0 \in \mathbb{R}^{N_S}$; external input $\vec \nu$ (arbitrary dimension); input mapping function $\mathbf{g}(\cdot)$}
\KwOut{Updated pre-CBS $\mathbf{q}$ after absorbing input and attention}

$\mathbf{q} \gets \mathbf{q} + \mathbf{g}(\vec \nu)$ \Comment{Incorporate external input}  

$\mathbf{q} \gets \mathbf{q} + 
\sum_{k=1}^{N_S} \mathbf{a}_k \, \mathrm{softmax} \Big(a_{0,k} + \sum_{\ell=1}^{n} a_{\ell,k} \, \mathrm{sig}(q_\ell) \Big)$ 
\Comment{Attention}  

\Return{$\mathbf{q}$}

\end{algorithm}

\begin{algorithm}[H]
\SetAlgoLined
\DontPrintSemicolon
\caption{Tensor Brain (TB) Generative Measurement}
\label{alg:cap-3NN}

\KwIn{pre-CBS $\mathbf{q} \in \mathbb{R}^n$; embedding matrix $\mathbf{A} \in \mathbb{R}^{n \times N_S}$; bias vector $\mathbf{a}_0 \in \mathbb{R}^{N_S}$; 
$0 \le \alpha \le 1$; $0 \le \beta \le 1$}
\KwOut{Sampled outcome $k$ and updated pre-CBS $\mathbf{q}$}

$k \sim \mathrm{softmax} \Big(a_{0,k} + \sum_{\ell=1}^{n} a_{\ell,k} \, \mathrm{sig}(q_\ell) \Big)$ \Comment{Generate outcome}  

$\mathbf{q} \gets \alpha \mathbf{q} + \beta \mathbf{a}_k$ \Comment{Update pre-CBS with outcome embedding}  

\Return{$k$, $\mathbf{q}$}

\end{algorithm}

Symbolic indices generated by {Algorithm~\ref{alg:cap-3NN}} may represent:  

\begin{itemize}
\item \textbf{Concepts:} entities, attributes, classes, or actions.  
\item \textbf{Predicates:} relations such as \textit{likes} or \textit{nextTo}.  
\item \textbf{Episodic markers:} temporal references, e.g., \textit{yesterday's lunch}. Even past thoughts and emotions.  
\item \textbf{Actions and decisions:} action of standing up, decision to drive home.
\end{itemize}

The mechanisms underlying the formation of symbolic indices, as well as their broader role in perception and memory, are beyond the scope of this paper. For further details, we refer to the foundational work on the Tensor Brain \citep{tresp2023tensor, tresp2025tensor}.

\section{Quantum Operations} 
\label{sec:qtheory}

Like classical probability theory, quantum theory is built upon a small set of axioms \citep{von2018mathematical}. A central concept is the \textbf{quantum state vector} 
$\vec{\psi} \in \mathbb{C}^N$, which describes a pure quantum state of dimension $N$. Postulate~1 of quantum mechanics states that a system is completely described by its quantum state vector.
In the limit $N \rightarrow \infty$, the formalism of Hilbert spaces becomes necessary. Quantum states satisfy the normalization constraint
\[
\| \vec{\psi} \|^2 = \vec{\psi}^\dagger \vec{\psi} = 1.
\]

Two fundamental operations in quantum theory are {time evolution} and {measurement}, both involving unitary operators. 
In a chosen basis, unitary operators are represented by unitary matrices $\mathbf{U} \in \mathbb{C}^{N \times N}$, satisfying
$
\mathbf{U}^\dagger \mathbf{U} = \mathbf{I},
$
where $\mathbf{U}^\dagger$ denotes the conjugate transpose, and $\mathbf{I}$ is the identity matrix. In what follows, we will use the terms ``matrix'' and ``operator'' interchangeably.

\subsection{Evolution Operator}

Let 
$\mathbf{U}^{\mathrm{evol}} \in \mathbb{C}^{N \times N}$ 
denote the time evolution operator, which is unitary. If an input vector $\vec \nu$ influences the evolution, we write $\mathbf{U}^{\mathrm{evol}}(\vec \nu)$.  

Postulate 2 of quantum mechanics states that evolution of a closed quantum system is described by a unitary transformation, as shown in 
{Algorithm~\ref{alg:cap-1Q}}.

\begin{algorithm}[H]
\SetAlgoLined
\DontPrintSemicolon
\caption{Quantum State Evolution}
\label{alg:cap-1Q}

\KwIn{Quantum state $\vec{\psi} \in \mathbb{C}^N$; unitary evolution operator $\mathbf{U}^{\mathrm{evol}} \in \mathbb{C}^{N \times N}$}
\KwOut{Quantum state $\vec{\psi}$ after applying the evolution operator}

$\vec{\psi} \gets \mathbf{U}^{\mathrm{evol}} \, \vec{\psi}$ \Comment{Update quantum state}  

\Return{$\vec{\psi}$}

\end{algorithm}

\subsection{Projective Measurements}   
\label{sec:pvm}

\subsubsection{PVM} 

Postulate 3 of quantum mechanics covers von Neumann  measurement processes described by projection operators, as described in this section. A more general formulation uses measurement operators, introduced in Section~\ref{sec:povmop}. 

A projection-valued measure (PVM) describes a projective quantum measurement, producing mutually exclusive, exhaustive, and repeatable outcomes.
It represents a standard projective quantum measurement. A PVM consists of a set of orthogonal projection operators 
$\{\mathbf{P}_k\}_{k \in K}$ acting on the Hilbert space $\mathbb{C}^N$, where each $\mathbf{P}_k$ corresponds to a possible measurement outcome, and $K$ is an indexing set.
Each projection operator is
 Hermitian 
(i.e., $\mathbf{P}_k^\dagger = \mathbf{P}_k$), pairwise orthogonal (i.e., $\mathbf{P}_j\mathbf{P}_k=0$ if $j\neq k$),
 and idempotent 
(i.e., $\mathbf{P}_k\mathbf{P}_k = \mathbf{P}_k$).
    Additionally, the set of projectors is complete $\sum_k \mathbf{P}_k =\mathbf{I}$.

If these criteria are satisfied, the probability of observing outcome $k$ is
\begin{equation} \label{eq:pvmmeas}
P(k) = \vec{\psi}^\dagger \mathbf{P}_k \vec{\psi}.
\end{equation}
After the measurement, the quantum state collapses to
\begin{equation} \label{eq:pvmresult}
\vec{\psi} \gets \frac{\mathbf{P}_k \vec{\psi}}{\sqrt{\vec{\psi}^\dagger \mathbf{P}_k \vec{\psi}}}.
\end{equation}

For an \emph{elementary} PVM, the projection operator is
\[
\mathbf{P}_k = \mathbf{u}_k \mathbf{u}_k^\dagger
\]
where $\mathbf{u}_k$ is the $k$-th column of $\mathbf{U}$. In this case, the probability of observing outcome $k$ is
\begin{equation} \label{eq:pkele}
P(k) = |\mathbf{u}_k^\dagger \vec{\psi}|^2
\end{equation}
and the post-measurement state becomes (up to a global phase)
\begin{equation} \label{eq:pkesus}
\vec{\psi} \gets \mathbf{u}_k.
\end{equation}

A characteristic feature of quantum systems is the probabilistic sampling of the measurement outcome and the replacement of the quantum state by $\mathbf{u}_k$. If the measurement is performed in the computational basis, then $\mathbf{U} = \mathbf{I}$, and an outcome $i$ results in
\[
\vec{\psi} \gets \mathbf{\tilde e}_i,
\] 
which corresponds to the well-known \emph{collapse of the wave function}. Here,  
$\mathbf{\tilde e}_i$ is a unit basis vector, i.e., a one-hot vector with the one at position $i$. 

In this paper, we refer to the computational basis as the $X$-basis.\footnote{In quantum systems this corresponding to a measurement in the Pauli-$Z$ basis \citep{nielsen2010quantum}}

\begin{algorithm}[H]
\SetAlgoLined
\DontPrintSemicolon
\caption{Projection-Valued Measurement (PVM)}
\label{alg:cap-3PVM}

\KwIn{Quantum state $\vec{\psi} \in \mathbb{C}^N$; unitary matrix $\mathbf{U} \in \mathbb{C}^{N \times N}$}
\KwOut{Sampled outcome $k$ and posterior quantum state $\vec{\psi}$}

Sample $k$ with probability $P(k) = |\mathbf{u}_k^\dagger \vec{\psi}|^2$ \Comment{Generate outcome}  

$\vec{\psi} \gets \mathbf{u}_k$ \Comment{State collapse}  

\Return{$k$, $\vec{\psi}$}

\end{algorithm}

\subsection{Generalized Measurements (POVMs)}

A {positive operator-valued measure (POVM)} provides the most general mathematical description of a quantum measurement.  
It captures effective measurements that account for noise, imperfections, or interactions with the environment.  
Before introducing POVMs formally, we first define the concept of a {density matrix}.

\subsubsection{Density Matrices}

A {density matrix} 
$\rho \in \mathbb{C}^{N \times N}$ 
is Hermitian, $\rho^\dagger = \rho$, positive semidefinite, and has unit trace, $\mathrm{tr}(\rho) = 1$.  
We use  $\mathrm{tr}$ to denote the trace operation,

A pure quantum state $\vec{\psi}$ can be represented as the density matrix
\[
\rho = \vec{\psi} \vec{\psi}^\dagger.
\]  
While state vectors, as discussed in the previous section, describe only pure states, 
 density matrices 
provide a complete and general description of quantum states, encompassing both pure and mixed cases, enabling analysis of subsystems, noise, and statistical mixtures — all of which are crucial in realistic quantum systems.

Time evolution in the density matrix formalism is given by
\begin{equation} \label{eq:rhoprop}
\rho \gets \mathbf{U}^{\mathrm{evol}} \, \rho \, (\mathbf{U}^{\mathrm{evol}})^\dagger.
\end{equation}

\subsubsection{POVM}
\label{sec:povmop}

A {positive operator-valued measure (POVM)} generalizes PVMs, allowing the description of more general measurements. A POVM consists of {measurement operators} $\mathbf{M}_k$ and {effect operators} (POVM elements) 
$
\mathbf{E}_k = \mathbf{M}_k^\dagger \mathbf{M}_k.
$  
The complete set $\{\mathbf{E}_k\}_{k \in K}$ defines the POVM.  

The requirements for POVMs are less restrictive than for PVMs:
     $\mathbf{E}_k$ are Hermitian (i.e., $\mathbf{E}_k^\dagger = \mathbf{E}_k$).   
     $\mathbf{E}_k$ are positive semidefinite (i.e., $\vec{\psi}^\dagger \mathbf{E}_k \vec{\psi} \ge 0$)  
     and are complete (i.e., $\sum_{k \in K} \mathbf{E}_k = \mathbf{I}$). The   
     $\mathbf{M}_k$  is defined as a matrix that factorizes the $\mathbf{E}_k$.

The  probability of observing outcome $k$ is (using the invariance of the trace under circular shifts)
\begin{equation} \label{eq:mpovm}
P(k) = \mathrm{tr}(\mathbf{E}_k \rho) = \mathrm{tr}(\mathbf{M}_k^\dagger \mathbf{M}_k \rho) = \mathrm{tr}(\mathbf{M}_k \rho \mathbf{M}_k^\dagger).
\end{equation}

After observing outcome $k$, the density matrix updates according to
\begin{equation} \label{eq:rhopropout}
\rho \gets \frac{\mathbf{M}_k \rho \mathbf{M}_k^\dagger}{\mathrm{tr}(\mathbf{M}_k \rho \mathbf{M}_k^\dagger)}.
\end{equation}

Unlike PVMs, POVMs are not necessarily idempotent, so repeated measurements may yield different outcomes. Moreover, POVMs allow for non-orthogonal measurement elements, enabling more general and sometimes more efficient descriptions. This includes the possibility of having either fewer or more valid outcomes than the dimension of the state space.  

For the special case of a PVM, we have
$
\mathbf{P}_k = \mathbf{M}_k = \mathbf{E}_k.
$

\subsubsection{HB-POVM}
\label{sec:hbo}

Given a unitary matrix $\mathbf{U}$, whose columns $\mathbf{u}_k$ form the measurement basis, we define a specific diagonal measurement operator
\begin{equation} \label{eq:mhb}
\mathbf{M}_k^{\mathrm{HB}} = \mathrm{diag}(\mathbf{u}_k),
\end{equation}
where $\mathrm{diag}(\mathbf{u}_k)$ denotes a diagonal matrix with the entries of the vector $\mathbf{u}_k$ on its diagonal.

The corresponding effect operator is
$
\mathbf{E}_k^{\mathrm{HB}} = \mathrm{diag}(\mathbf{b}_k),
$ 
which satisfies the POVM conditions. The set of operators $\{\mathbf{E}_k^{\mathrm{HB}}\}$ defines the \textit{Heisenberg--Bayes POVM (HB-POVM)}. Our naming is due to the fact that there is a special case where probabilistic quantum and Bayes formula give the same result (See Section~\ref{sec:canonicalclassical}).

Suppose the system is in a pure state $\vec{\psi}$ with density matrix
$
\rho = \vec{\psi} \vec{\psi}^\dagger.
$  
Define the probabilistic prior 
$p_i = |\psi_i|^2,$ and  $\mathbf{p} = (p_1, \ldots, p_N)^\top.$
Then, the probability of observing outcome $k$ under the HB-POVM is
\begin{equation} \label{eq:pkhbo}
P(k) = \mathrm{tr}\big(\mathbf{M}^{\mathrm{HB}}_k \, \vec{\psi} \vec{\psi}^\dagger \, (\mathbf{M}^{\mathrm{HB}}_k)^\dagger \big) 
= \mathrm{tr}\big( (\mathbf{u}_k \circ \vec{\psi}) (\mathbf{u}_k \circ \vec{\psi})^\dagger \big) 
= \mathbf{b}_k^\top \mathbf{p},
\end{equation}
where $\circ$ denotes the Hadamard (elementwise) product.

After outcome $k$, the posterior quantum state becomes
\begin{equation} \label{eq:suppovm}
\vec{\psi} \gets \frac{\mathbf{M}^{\mathrm{HB}}_k \, \vec{\psi}}{\sqrt{P(k)}} 
= \frac{\mathbf{u}_k \circ \vec{\psi}}{\sqrt{P(k)}}.
\end{equation}

The corresponding posterior density matrix is
$
\rho \gets \vec{\psi} \vec{\psi}^\dagger,
$
where $\vec{\psi}$ is the updated quantum state.

Here, $\mathbf{b}_k$ is the $k$-th column of the unitary stochastic matrix $\mathbf{B} \in [0,1]^{N \times N}$, defined by
\[
b_{i,k} = |u_{i,k}|^2, \quad u_{i,k} = (\mathbf{U})_{i,k}.
\]  
A unitary-stochastic matrix has the property that it is doubly stochastic, which means that  both its rows and columns are valid probability vectors.

A distinctive feature of the HB-POVM is that the quantum state prior to measurement is not completely discarded; instead, it is combined with the measurement vector $\mathbf{u}_k$ via a Hadamard (elementwise) product.  
This property provides a natural explanation for the skip connection observed in the Tensor Brain (TB) framework (see Section~\ref{sec:sconnz2}).  

Algorithm~\ref{alg:cap-3POVM} implements the HB-POVM.

\begin{algorithm}[H]
\SetAlgoLined
\DontPrintSemicolon
\caption{Heisenberg--Bayes POVM (HB-POVM)}
\label{alg:cap-3POVM}

\KwIn{Quantum state $\vec{\psi} \in \mathbb{C}^N$; unitary matrix $\mathbf{U}^{\mathrm{evol}} \in \mathbb{C}^{N \times N}$}
\KwOut{Measurement outcome $k$ and posterior quantum state $\vec{\psi}$}

Form vector $\mathbf{p}$ with $p_i = |\psi_i|^2$\;  

Form matrix $\mathbf{B}$ with $b_{i,k} = |u_{i,k}|^2$ \Comment{Unitary-stochastic matrix}

Sample $k$ with probability $P(k) = \mathbf{b}_k^\top \mathbf{p}$ \Comment{Measurement outcome}  

$\vec{\psi} \gets {\mathbf{u}_k \circ \vec{\psi}} / {\sqrt{\mathbf{b}_k^\top \mathbf{p}}}$ \Comment{Posterior state update}

\Return{$k$, $\vec{\psi}$}

\end{algorithm}

\subsection{Postselection}
\label{sec:posts}

Often, only a subset of measurement outcomes is relevant. {Postselection} restricts the results to this subset, 
\textit{e.g.}, by discarding or terminating experiments with outcomes that are not retained.  
Postselection has been studied in quantum information in the context of complexity theory~\citep{aaronson2005quantum}.

Define an indicator vector 
\[
\mathbf{z} \in \{0,1\}^N, \quad z_k = 1 \text{ if outcome $k$ is retained, otherwise } 0.
\]  
The renormalized probability distribution over the postselected outcomes is
\begin{equation} \label{eq:pspvm}
P(k) = \frac{z_k \, |\mathbf{u}_k^\dagger \vec{\psi}|^2}{\sum_{k: z_k = 1} |\mathbf{u}_k^\dagger \vec{\psi}|^2}.
\end{equation}

Outcomes that are not postselected are assigned probability $0$, while the probabilities of the retained outcomes are renormalized to sum to 1.

For the HB-POVM, the renormalized probability becomes
\begin{equation} \label{eq:povmmea}
P(k)  =   \frac{z_k  \mathbf{b}_k^\top  \mathbf{p}}
{\sum_{k: z_k =1} \mathbf{b}_k^\top  \mathbf{p}} .
\end{equation}

\subsection{Qubits}
\label{sec:qubitreprsup}

For $N = 2^n$, let the index $i$ have the binary expansion
\[
i(i_1, \ldots, i_n) = 1 + \sum_{\ell=1}^{n} i_\ell 2^{\,\ell-1}, \quad i_\ell \in \{0,1\}.
\]  
Here, $i(i_1, \ldots, i_n)$ indicates that $i$ is a function of the bits $i_1, \ldots, i_n$. The map is invertible and we can write $i_\ell (i), \ell = 1, \ldots, n$.

Then, the quantum state can be expressed as
\[
\vec{\psi} = \sum_{i=1}^N \psi_i \, \mathbf{\tilde e}_i
= \sum_{i_1, \ldots, i_n} \psi_{i(i_1, \ldots, i_n)} \, \mathbf{e}_{i_1} \otimes \cdots \otimes \mathbf{e}_{i_n},
\]
where 
\[
\mathbf{e}_0 = \begin{pmatrix} 1 \\ 0 \end{pmatrix}, \quad
\mathbf{e}_1 = \begin{pmatrix} 0 \\ 1 \end{pmatrix},
\]
and $\mathbf{\tilde e}_i$ is the $N$-dimensional standard basis vector with a $1$ in the $i$-th position and zeros elsewhere.  

The notation $\psi_{i_1, \ldots, i_n}$ indicates that the quantum state can be written as a tensor. 
This mapping from a vector to an $n$-mode tensor is called {tensorization} \citep{hackbusch2012tensor}.

\paragraph{Measurements.}  
Let $i$ be the outcome of a
a PVM measurement in the computational basis. Then 
\[
\vec{\psi} \leftarrow
\mathbf{e}_{i_1(i)} \otimes \cdots \otimes \mathbf{e}_{i_n(i)},
\]
In a physical realization, we can identify $\mathbf{e}_{i_\ell(i)}$ as the state of physical qbubit $\ell$.



It is also possible to perform measurements on individual qubits.  For example, the probability of measuring the first qubit with outcome $1$ is
\[
P(i_1 = 1) = \sum_{i_2, \ldots, i_n} |\psi_{i_1=1, i_2, \ldots, i_n}|^2.
\]

After such a measurement, the quantum state updates to
\[
\vec{\psi} \gets \frac{1}{\sqrt{P(i_1 = 1)}} \sum_{i_2, \ldots, i_n} \psi_{i(i_1=1, i_2, \ldots, i_n)} \;
\mathbf{e}_{i_1=1} \otimes \mathbf{e}_{i_2} \otimes \cdots \otimes \mathbf{e}_{i_n}.
\]
Thus, qubits $2, \ldots, n$ remain in superposition.
This instantaneous update of the quantum state underlies the famous EPR paradox.  

The measurement just described is a  \textit{partial measurement}, also referred to as a \textit{degenerate measurement} in other contexts. Both cases are discussed in Appendix~\ref{sec:degpam}.

\paragraph{Entanglement.}  
If the quantum state factorizes as
\[
\psi_{i(i_1, i_2, \ldots, i_n)} = \prod_{\ell=1}^n \psi_{i_\ell},
\]
then the qubits are separable. Otherwise, the qubits are {entangled}.  

From the perspective of tensorization, entanglement arises naturally as a property of the tensor representation.  
In contrast, starting with physical qubits, entanglement often appears more mysterious.

\paragraph{Evolution.}  
In the qubit representation, the evolution operator can be expressed as
\begin{equation} \label{eq:qcpower}
\psi_{j(j_1, \ldots, j_n)} \gets \sum_{i_1, \ldots, i_n} 
u^{\mathrm{evol}}_{j(j_1, \ldots, j_n), \, i(i_1, \ldots, i_n)} \; 
\psi_{i(i_1, \ldots, i_n)} \qquad \forall \{j_1, \ldots, j_n\}.
\end{equation}
This operation is known as a {tensor contraction}. We defined
$u^{\mathrm{evol}}_{j,i} = (\mathbf{U}^{\mathrm{evol}})_{j, i}$.

\subsubsection{Quantum Computer}
\label{sec:qcc}

In a quantum computer, the evolution operator is implemented as a {quantum circuit}, consisting of a sequence of gates acting on one or a few qubits at a time.  

For example, a two-qubit gate acting on qubits 1 and 2, while leaving the remaining qubits unchanged, corresponds to a unitary matrix with elements
\[
u_{j(j_1, \ldots, j_n), \, i(i_1, \ldots, i_n)} = u_{j_1, j_2, i_1, i_2} \prod_{\ell=3}^{n} \delta_{j_\ell, i_\ell},
\]
where $u_{j_1, j_2, i_1, i_2}$ is an entry of a $4 \times 4$ unitary matrix, and the Kronecker deltas enforce that untouched qubits pass through unchanged,
i.e., if qubit $\ell \ge 3$ is in state $\mathbf{e}_{i_\ell}$ it will be in the same 
state $\mathbf{e}_{i_\ell}$, after the application of the quantum gate.  

If the circuit consists of $L$ gates, the overall evolution operator is
\begin{equation} \label{eq:qcp}
\mathbf{U} = \mathbf{U}^{(L)} \cdots \mathbf{U}^{(2)} \mathbf{U}^{(1)}.
\end{equation}
$\mathbf{U}^{(\ell)} $ described the operation of quantum gate $\ell$.  
The factorization of $\mathbf{U}$ into a quantum circuit is neither trivial nor  unique. 

Quantum computers can solve certain problems efficiently. A canonical example is the quantum Fourier transform (QFT) on an $N$-dimensional quantum state, which can be performed on $n = \log_2 N$ qubits in $\mathcal{O}(n^2)$ operations. A classical FFT requires $\mathcal{O}(Nn)$ operations.

\subsubsection{Qubit Measurements}
\label{sec:one-sided}

With a measurement basis defined by 
$\mathbf{U}$, tensorization can be performed as a 
 two-sided tensorization $u_{i(i_1, \ldots, i_n), k(k_1, \ldots, k_n)}$
and a one-sided tensorization $u_{i(i_1, \ldots, i_n), k}$. 
Two-sided tensorization is addressed in Section~\ref{sec:distrind}.


\section{Probabilistic Quantum
}
\label{sec:probq}

We now revisit the quantum system described earlier, with the key modification that we impose \emph{well-defined measurements}.  
This change allows the system to be analyzed within a probabilistic framework: the outcomes of measurements become random variables whose distributions are fully determined by the initial quantum state, the evolution protocol, and the chosen measurement operators.  

Since the uncertainty still originates from the underlying quantum system, we refer to this approach as \emph{probabilistic quantum}.

\subsection{Ignorant Measurement}
\label{sec:uuoo}

An \emph{ignorant measurement} is a measurement performed without the observer recording or learning the outcome  
\cite[p.~159]{wilde2013quantum}; see also \cite[p.~100]{nielsen2010quantum}.  
In quantum mechanics, however, the act of measurement—even without observing the result—inevitably disturbs the system.  
The term \emph{nonselective measurement} emphasizes that this disturbance is not merely \textit{epistemic} (reflecting only the observer’s ignorance) but also \textit{ontic} (affecting the physical state of the system itself).  

Formally, applying an ignorant measurement to a density matrix $\rho$ results in
\begin{equation} \label{eq:ignm}
\rho \;\leftarrow\; \sum_i \mathbf{M}_i \, \rho \, \mathbf{M}_i^\dagger .
\end{equation}


For an ignorant measurement in the $X$-basis, with measurement operators $\mathbf{M}_i = \mathbf{e}_i \mathbf{e}_i^\top$, the diagonal entries of $\rho$ are preserved while all off-diagonal terms vanish.  
In particular, if the system is initially in a pure state $\vec{\psi}$, then
\[
\rho \;\leftarrow\; \mathrm{diag}(\mathbf{p}), 
\quad \text{with } p_i = |\psi_i|^2 .
\]
Thus, an unobserved $X$-measurement transforms the density matrix into a diagonal form whose entries are exactly the outcome probabilities $p_i$.  

We define the vector of diagonal entries $\mathbf{p}$ as the \emph{probabilistic state}.  
This representation allows us to model the system entirely within a probabilistic framework,  
without explicit reference to the underlying wavefunction or unitary operators.  

In this probabilistic formulation, it is useful to distinguish two random variables.  
First, the \emph{state variable} $X$, with  
  \[
  P(X = i \mid \cdot) \equiv p_i 
  \]
  representing the latent distribution over basis states, conditioned on all prior evolution and measurement steps.  
Second,  \emph{observed variable} $Y$, with  
  \[
  P(Y = k \mid \cdot) \equiv P(k) 
  \]
  corresponding to the actual measurement outcome in the chosen protocol.

In our setting, we assume that an ignorant $X$-measurement occurs after each evolution step and after each $Y$-measurement.



\subsection{Probabilistic State Evolution}

Suppose the quantum system is in a probabilistic state $\mathbf{p}$, e.g., after an ignorant measurement.  
After applying the evolution operator and using Equation~\ref{eq:rhoprop}, the density matrix becomes
\[
\rho \;\leftarrow\; 
\sum_i p_i \, \mathbf{u}^{\text{evol}}_i (\mathbf{u}^{\text{evol}}_i)^\dagger 
\]
that is, a mixed state formed from pure states $\mathbf{u}^{\text{evol}}_i$ weighted by the prior probabilities $p_i$.  

Another ignorant measurement then removes all off-diagonal terms, yielding
\begin{equation} \label{eq:pproa}
 \mathbf{p} \;\leftarrow\; \mathbf{B}^{\text{evol}} \mathbf{p},  
\end{equation}
where $\mathbf{B}^{\text{evol}}$ is the unitary-stochastic matrix induced by the unitary $\mathbf{U}^{\text{evol}}$, i.e., 
$(\mathbf{B}^{\text{evol}})_{j, i} = |(\mathbf{U}^{\text{evol}})_{j, i}|^2$ .  
Thus, the evolution of $\rho$ can be equivalently described as a linear transformation of the probabilistic state $\mathbf{p}$.  
This yields a fully probabilistic formulation of quantum dynamics (see Algorithm~\ref{alg:cap-1E}).

\begin{remark}
Equation~\eqref{eq:pproa} defines a \emph{Markovian update rule}:  
the probabilistic state $\mathbf{p}$ evolves by multiplication with the stochastic matrix $\mathbf{B}^{\text{evol}}$.  
In this way, the underlying unitary quantum dynamics reduce to a classical Markov process acting on the latent state variable $X$.
\end{remark}


\begin{algorithm}[H]
 \SetAlgoLined
 \DontPrintSemicolon
 \caption{Probabilistic State Evolution}
 \label{alg:cap-1E}

 \KwIn{Probabilistic state $\mathbf{p} \in [0,1]^N$; 
       stochastic matrix $\mathbf{B}^{\text{evol}} \in [0,1]^{N \times N}$ (unitary-stochastic).}
 \KwOut{Updated probabilistic state $\mathbf{p}$.}

 $\mathbf{p} \;\leftarrow\; \mathbf{B}^{\text{evol}} \mathbf{p}$ \Comment{Markovian update} \; 

 \Return{$\mathbf{p}$}\;
\end{algorithm}

\subsection{Probabilistic PVM}
\label{sec:probpvm}


Suppose the system is in a probabilistic state $\mathbf{p}$ (e.g., following an $X$-measurement with an unknown outcome).  
We now consider a projective measurement in the $Y$-basis, defined by the unitary matrix $\mathbf{U}$.  

\paragraph{Outcome probability.}  
The probability of obtaining outcome $k$ is
\begin{equation} \label{eq:propk}
P(k) \;=\; \mathrm{tr}\!\left( \mathbf{M}_k \rho \mathbf{M}_k^\dagger \right) 
= \mathrm{tr}\!\left( \mathbf{u}_k \mathbf{u}_k^\dagger \rho \mathbf{u}_k \mathbf{u}_k^\dagger \right)
\end{equation}
$$
= 
\left(\sum_i p_i |\mathbf{u}_{i, k}|^2 \right)\;  \mathrm{tr}\! \left( \mathbf{u}_k   \mathbf{u}_k^\dagger 
\right)
= \mathbf{b}_k^\top \mathbf{p},
$$
where $\mathbf{u}_k$ is the $k$-th column of $\mathbf{U}$, and $\mathbf{b}_k$ is the $k$-th column of the associated unitary-stochastic matrix $\mathbf{B}$ defined in Section~\ref{sec:hbo}.

\paragraph{Posterior state.}  
If outcome $k$ is observed, the posterior state is
\[
\rho \;\leftarrow\; 
\frac{1}{P(k)} \left(\sum_i p_i |\mathbf{u}_{i, k}|^2\right)\;  \mathbf{u}_k   \mathbf{u}_k^\dagger 
\;=\; \mathbf{u}_k \mathbf{u}_k^\dagger .
\]
Thus, as expected, the system collapses to the pure state 
$\vec \psi \leftarrow \mathbf{u}_k$.  

\paragraph{Ignorant measurement of the posterior state.}  
If an ignorant $X$-measurement is subsequently performed, all off-diagonal terms vanish, leaving the probabilistic state
\begin{equation} \label{eq:pvmpost}
\mathbf{p} \;\leftarrow\; \mathbf{b}_k .
\end{equation}

\begin{algorithm}[H]
 \SetAlgoLined
   \DontPrintSemicolon
\caption{Probabilistic     PVM}
\label {alg:cap-3PVM }
\KwIn{Probabilistic  state $\mathbf{p} \in [0, 1]^N $; unitary-stochastic  matrix  $\mathbf{B}^{\text{evol}}  \in [0,1]^{N \times N} $}
\KwOut{Outcome $k$ and  posterior probabilistic state $\mathbf{p}$}

  Sample $k$ with  $P(k) =  \mathbf{b}_k^\top  \mathbf{p}$ \Comment{Outcome}
  
 
 $\mathbf{p}\leftarrow  \mathbf{b}_k$  \Comment{State update}
 
\Return{$k$, $\mathbf{p}$}\;
\end{algorithm}

\subsection{Probabilistic HB-POVM}
\label{sec:hbpovvm}

\paragraph{Outcome probability.}  
Recall the operators $\mathbf{M}^{\textit{HB}}_k$ introduced in Section~\ref{sec:hbo}.  
For a probabilistic state $\mathbf{p}$, the probability of outcome $k$ is  
\[
P(k) \;=\; \mathrm{tr}\!\left(\rho \, \mathbf{E}^{\textit{HB}}_k \right) 
= \mathrm{tr}\!\left(\mathbf{M}^{\textit{HB}}_k \, \mathrm{diag}(\mathbf{p}) \, (\mathbf{M}^{\textit{HB}}_k)^\dagger \right) 
= \mathrm{tr}\!\left(\sum_i p_i \, |u_{i,k}|^2  \right)
= \mathbf{b}_k^\top \mathbf{p}.
\]

\paragraph{Posterior state.}  
According to Equation~\ref{eq:rhopropout}, the post-measurement density matrix 
\[
\rho \;\leftarrow\; \frac{1}{P(k)} \, \mathrm{diag}\!\left(\mathbf{b}_k \circ \mathbf{p}\right).
\]

\paragraph{Ignorant measurement of the posterior state.}  
Applying an ignorant $X$-measurement eliminates off-diagonal terms and yields the updated probabilistic state  
\begin{equation} \label{eq:BF1}
\mathbf{p} \;\leftarrow\; 
\frac{\mathbf{b}_k \circ \mathbf{p}}{\mathbf{b}_k^\top \mathbf{p}} .
\end{equation}

\begin{algorithm}[H]
 \SetAlgoLined
   \DontPrintSemicolon
\caption{Probabilistic  HB-POVM}
\label {alg:cap-3EPOVM}
\KwIn{Probabilistic  state $\mathbf{p} \in [0, 1]^N $; unitary-stochastic  matrix  $\mathbf{B}^{\text{evol}}  \in [0,1]^{N \times N}  $}
\KwOut{Outcome $k$ and  posterior probabilistic state $\mathbf{p}$}

  Sample $k$ with  $P(k) =  \mathbf{b}_k^\top \mathbf{p}$ \Comment{Outcome}

 $\mathbf{p} \leftarrow (  \mathbf{b}_k \circ \mathbf{p} ) / 
 (\mathbf{b}_k^\top \mathbf{p})$ \Comment{State update}
 
\Return{$k$, $\mathbf{p}  $}\;
\end{algorithm}

\subsection{Postselection  for Probabilistc Quantum}
\label{sec:probposts}

We now extend the probabilistic framework to incorporate \emph{postselection}, as introduced in Section~\ref{sec:posts}.  
The probability of observing outcome $k$ under postselection remains given by Equation~\ref{eq:povmmea}, for both the probabilistic PVM and the probabilistic HB-POVM.  
Importantly, postselection does not alter the update rules for the probabilistic state:  
\begin{itemize}
    \item In the probabilistic PVM, the posterior is given by Equation~\ref{eq:pvmpost}.  
    \item In the probabilistic HB-POVM, the posterior is given by Equation~\ref{eq:BF1}.  
\end{itemize}

\begin{figure}
    \centering
    \begin{subfigure}[b]{0.5\textwidth}
        \centering
        \resizebox{\linewidth}{!}{
        \begin{tikzpicture}[
        xstate/.style={circle, draw=lightgreen, very thick, minimum size=0.8cm, fill=white},
        ystate/.style={circle, draw=darkred, very thick, minimum size=0.8cm, fill=white},
        arrow/.style={->, thick, darkblue},
        dotted_arrow/.style={->, thick, dotted, darkblue},
        evolution_arrow/.style={->, thick, black},
        label/.style={font=\large\bfseries},
        sublabel/.style={font=\normalsize},
        postselect/.style={circle, draw=darkblue, thick, minimum size=0.8cm, fill=lightblue!20, inner sep =1pt}]

        \node[xstate] (X-1) at (-4, 0) {};
        \node[xstate] (X) at (-2, 0) {$X$};
        \node[ystate] (Y) at (0, 0) {$Y$};
        \node[xstate] (X_prime) at (2, 0) {$X'$};
        \node[xstate] (X_prime+1) at (4, 0) {};
        
        \draw[evolution_arrow] (X-1) -- (X);
        \draw[evolution_arrow] (X) -- (Y);
        \draw[evolution_arrow] (Y) -- (X_prime);
        \draw[evolution_arrow] (X_prime) -- (X_prime+1);
        
        \node[sublabel] at (-3, 0.4) {$B^{\text{evol}}$};
        \node[sublabel] at (-1, 0.4) {$B^T$};
        \node[sublabel] at (1, 0.4) {$B$};
        \node[sublabel] at (3, 0.4) {$B^{\text{evol}}$};
        
        \draw[dotted_arrow, bend right=45] (X) to node[midway, below, font=\normalsize] {HB-POVM} (X_prime);
        
        \node[postselect] (Z) at (0, 2) {\small $Z=1$};
        \node[sublabel] at (2, 2) {Postselection};
        
        \draw[dotted_arrow] (0, 0.5) -- (Z);
        \end{tikzpicture}
        }
        \caption{Probabilistic PVM}
        \label{fig-prob-pvmF}
    \end{subfigure}

    \hspace{5pt}

    \begin{subfigure}[b]{0.34\textwidth}
        \centering
        \resizebox{\linewidth}{!}{
        \begin{tikzpicture}[
        xstate/.style={circle, draw=lightgreen, very thick, minimum size=0.8cm, fill=white},
        ystate/.style={circle, draw=darkred, very thick, minimum size=0.8cm, fill=white},
        arrow/.style={->, thick, darkblue},
        dotted_arrow/.style={->, thick, dotted, darkblue},
        evolution_arrow/.style={->, thick, black},
        label/.style={font=\large\bfseries},
        sublabel/.style={font=\normalsize},
        postselect/.style={circle, draw=darkblue, thick, minimum size=0.8cm, fill=lightblue!20, inner sep=1pt}
        ]
        \node[xstate] (X-1) at (-2, 0) {};
        \node[xstate] (X) at (0, 0) {$X$};
        \node[xstate] (X+1) at (2, 0) {};
        
        \draw[evolution_arrow] (X-1) -- (X);
        \draw[evolution_arrow] (X) -- (X+1);
        
        \node[sublabel] at (-1, 0.3) {$B^{\text{evol}}$};
        \node[sublabel] at (1, 0.3) {$B^{\text{evol}}$};
        
        \node[ystate] (Y) at (0, 2) {$Y$};
        \draw[evolution_arrow] (X) -- (Y);
        \node[sublabel] at (0.4, 1) {$B^T$};

        \node[postselect] (Z) at (2, 2) {\small $Z=1$};
        \node[sublabel] at (2, 2.8) {\small Postselection};
        \draw[dotted_arrow] (Y) -- (Z);
        
        \end{tikzpicture}
        }
        \caption{Prob. HB-POVM}
        \label{fig-prob-hb-povmF}
    \end{subfigure}
    \hfill
    \begin{subfigure}[b]{0.34\textwidth}
        \centering
        \resizebox{\linewidth}{!}{
        \begin{tikzpicture}[
        postselect/.style={circle, draw=darkblue, thick, minimum size=0.8cm, fill=lightblue!20, inner sep=1pt},
        xstate/.style={circle, draw=lightgreen, very thick, minimum size=0.8cm, fill=white},
        ystate/.style={circle, draw=darkred, very thick, minimum size=0.8cm, fill=white},
        arrow/.style={->, thick, darkblue},
        dotted_arrow/.style={->, thick, dotted, darkblue},
        evolution_arrow/.style={->, thick, black},
        label/.style={font=\large\bfseries},
        sublabel/.style={font=\normalsize}
        ]
        
        \node[xstate] (X-1) at (-2, 0) {};
        \node[xstate] (X) at (0, 0) {$X$};
        \node[xstate] (X+1) at (2, 0) {};
        
        \draw[evolution_arrow] (X-1) -- (X);
        \draw[evolution_arrow] (X) -- (X+1);
        
        \node[sublabel] at (-1, 0.3) {$B^{\text{evol}}$};
        \node[sublabel] at (1, 0.3) {$B^{\text{evol}}$};
        
        \node[ystate] (Y) at (0, 2) {$Y$};
        \draw[evolution_arrow] (X) -- (Y);
        \node[sublabel] at (0.4, 1) {$\bar{B}^T$};

        \node[postselect] (Z) at (2, 2) {\small $Z=1$};
        \node[sublabel] at (2, 2.8) {\small Postselection};
        \draw[dotted_arrow] (Z) -- (Y);
        
        \end{tikzpicture}
        }
        \caption{gHMM}
        \label{fig-prob-gmm}
    
    \end{subfigure}
    \hfill
    \begin{subfigure}[b]{0.3\textwidth}
        \centering
        \resizebox{\linewidth}{!}{
        \begin{tikzpicture}[
        postselect/.style={circle, draw=darkblue, thick, minimum size=0.8cm, fill=lightblue!20, inner sep=1pt},
        xstate/.style={circle, draw=lightgreen, very thick, minimum size=0.8cm, fill=white},
        ystate/.style={circle, draw=darkred, very thick, minimum size=0.8cm, fill=white},
        arrow/.style={->, thick, darkblue},
        dotted_arrow/.style={->, thick, dotted, darkblue},
        evolution_arrow/.style={->, thick, black},
        label/.style={font=\large\bfseries},
        sublabel/.style={font=\normalsize}
        ]
        
        \node[xstate] (X-1) at (-2, 0) {};
        \node[xstate] (X) at (0, 0) {$X$};
        \node[xstate] (X+1) at (2, 0) {};
        
        \draw[evolution_arrow] (X-1) -- (X);
        \draw[evolution_arrow] (X) -- (X+1);
        
        \node[sublabel] at (-1, 0.3) {$B^{\text{evol}}$};
        \node[sublabel] at (1, 0.3) {$B^{\text{evol}}$};
        
        \node[ystate] (Y) at (0, 2) {$Y$};
        \draw[evolution_arrow] (X) -- (Y);
        \node[sublabel] at (0.4, 1) {$\bar{B}^T$};
        
        \end{tikzpicture}
        }
        \caption{gHMM w. $Z=1$}
        \label{fig-prob-gmm2}
    \end{subfigure}

    \caption{
\textit{Top:} The probabilistic PVM.  
The random variable $Y$ is part of the Markov chain defined by the random variables $X, X', \ldots$. $X$ and $X'$ are unmeasured and remain in  probabilistic states (indicated by green circles).  
In postselection, the binary variable $Z=1$ filters the selected outcomes.  
\textit{Bottom left:} The probabilistic HB-POVM represented as a valid probabilistic model.  
Under postselection, dependencies exist from $X$ to $Y$ and from $Y$ to $Z$.  \textit{Bottom center and right:} The generalized HMM (gHMM).  
Figure (c) shows the collider structure, while (d) illustrates the resulting model.  
}
    \label{fig:PVM}
\end{figure}
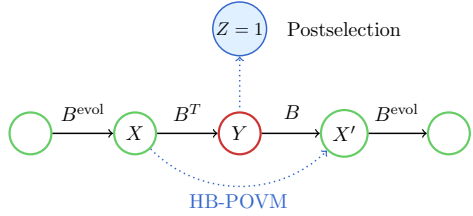
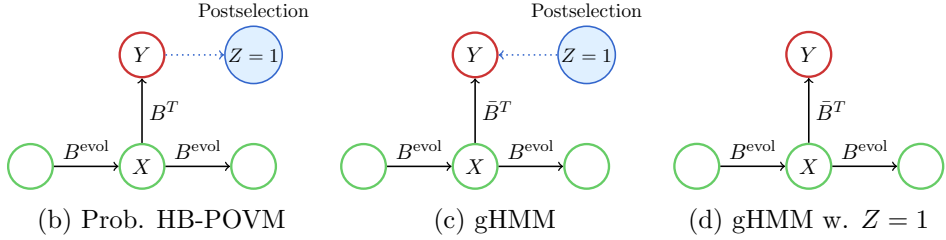

Figure~\ref{fig:PVM} illustrates the resulting dependency structure.  
A key point is that the random variable $Z$ is a child of $Y$, where $P(Z =1 | Y=k) = z_k$.  
This ensures that the probabilistic model remains consistent and preserves the correct causal interpretation.

\subsection{The Canonical Classical System}
\label{sec:canonicalclassical}

The distinctive feature of quantum theory is that probabilistic sampling is intrinsic to the system itself, i.e., it is \emph{ontic}. 
Even in a perfectly isolated system, measurement outcomes remain uncertain.  
In contrast, the classical world is fundamentally deterministic---for example, the predictable stability of planetary orbits. 
Classical measurements do not disturb the underlying state.  

In classical settings, probability arises only from an observer’s lack of information, i.e., \emph{epistemic uncertainty}, 
which can be captured by a probabilistic model. Such models are widely used in simulation, for instance in generative AI.  
Of particular relevance to our discussion is the \emph{generative hidden Markov model} (gHMM).

 \subsubsection{The Generative Hidden Markov Model (gHMM)}

Consider a generative hidden Markov model (gHMM) with state transition matrix $\mathbf{B}^{\text{evol}}$ and observation likelihood
\[
P(Y = k \mid X = i) \;=\; b_{i,k}.
\]
This identification is consistent because the unitary-stochastic matrix $\mathbf{B}$ has both rows and columns that form probability vectors.  
Consequently, Equation~\ref{eq:BF1} coincides exactly with Bayes’ rule for the gHMM.  

Hence, the HB-POVM induces a quantum measurement process that directly mirrors Bayesian updating of probabilistic states.  
As illustrated in Figure~\ref{fig:PVM} (bottom left), in the absence of postselection, the probabilistic HB-POVM and the Bayes observer equations are identical.  

This establishes an important connection:  
\textit{a probabilistic quantum system can be viewed as performing Bayesian updates on a classical probabilistic state.}

{ This equivalence relies on the doubly stochastic property of the matrix $\mathbf{B}$.  
Without this property, inference in the probabilistic HB-POVM and the gHMM can diverge.  
The same holds when postselection is applied.}  

In the postselection scenario, the random variable $Z$ acts as a parent of $Y$, making $Y$ a collider.  
The classical Bayes model is illustrated in Figure~\ref{fig:PVM} (bottom right).  
Given $Z=1$, the likelihood must be properly normalized, which is accomplished by defining
\[
\bar b_{i,k} = \frac{z_k \, b_{i,k}}{\sum_j z_j \, b_{i,j}}.
\]
We then have
\begin{equation} \label{eq:bmeass}
P(k) = \mathbf{\bar b}_k^\top \mathbf{p}.
\end{equation}
For a given observation $k$, the posterior probabilistic state is updated as
\begin{equation} \label{eq:bmeass987}
\mathbf{p} \;\leftarrow\; \frac{1}{P(k)} \, \mathbf{\bar b}_k \circ \mathbf{p}.
\end{equation}

This is the standard Bayes’ update formula, but with a modified conditional probability due to postselection.  
In contrast, in the HB-POVM framework, one would use $\mathbf{b}_k$ instead of $\mathbf{\bar b}_k$ in the above two equations.  
Although the difference may be small, it is significant, when the states of the random variables are tensorized and described by the states of pro-bits (Section~\ref{sec_probitr}) or neurons (Section~\ref{sec:fact}).

\subsubsection{Interpreting  $\mathbf{B}$ and $\mathbf{\bar B}$}

For both the probabilistic PVM and the probabilistic HB-POVM, the column-normalized version of $\mathbf{B}$ must be specified for the postselected outcomes.  
These columns can be interpreted as \textit{posterior} probabilities, such that for all postselected outcomes $k$,
\[
b_{i, k} \equiv P(X=i \mid Y=k).
\] 
In the case of the HB-POVM, this interpretation assumes a uniform prior $P(X=i) = 1/N, \forall i $.  

In contrast, for probabilistic inference in the gHMM, only the row-normalized version $\mathbf{\bar B}$ (restricted to the postselected outcomes) has a direct probabilistic interpretation:  
\[
\bar b_{i, k} \equiv P(Y=k \mid X=i).
\]  
Here, the conditional \textit{likelihood} is specified, rather than the posterior.

\subsection{A Conjecture}
\label{sec:conj}

We have seen that the outcomes of  well-defined measurements follow a probability distribution. 
We formulate the following conjecture: 

\begin{conjecture}

Given an initial quantum state, a fixed sequence (i.e.,  protocol) 
of evolution operations and measurements, and an external inputs, 
the resulting measurement outcomes can be described probabilistically.
\end{conjecture}

In Section~\ref{sec:qinf}, we will explain why this conjecture does not conflict with the distinct interference effects observed in quantum systems.

\subsection{Pro-bit Representation}
\label{sec_probitr}

\subsubsection{Pro-bit Representation}
\label{sec:onequbit}

The construction is fully analogous to the quantum case.  
Using the index notation introduced in Section~\ref{sec:qubitreprsup}, the probability distribution can be expressed as
\[
p_i = p_{i(i_1, \ldots, i_n)} = p_{i_1, \ldots, i_n}
\]
where $i(i_1, \ldots, i_n)$ is defined as before (Section~\ref{sec:qubitreprsup}). 

The probabilistic state vector
\[
\mathbf{p} = \sum_{i=1}^N p_i \, \mathbf{\tilde e}_i
\]
can equivalently be written in tensor-product form as
\[
{\mathbf{p}} = \sum_{i(i_1, \ldots, i_n)} p_{i(i_1, \ldots, i_n)} \; 
   \mathbf{e}_{i_1} \otimes \cdots \otimes \mathbf{e}_{i_n}.
\]

\paragraph{Measurements.}  
Pro-bits can be treated as measurable quantities.  
A \emph{probabilistic PVM} corresponds to measuring all pro-bits simultaneously, in which case each pro-bit collapses into either state $\mathbf{e}_0$ or $\mathbf{e}_1$.  
Alternatively, one can perform a \emph{partial measurement}.  
For example, measuring only the first pro-bit yields, by marginalization, 
\[
P(i_1 = 1) = \sum_{i_2, \ldots, i_n} p_{i_1=1, i_2, \ldots, i_n}.
\]
By conditioning, the probabilistic state is updated to
\[
{\mathbf{p}} \;\;\leftarrow\;\; 
\frac{1}{P(i_1=1)} \sum_{i_2, \ldots, i_n} p_{i_1=1, i_2, \ldots, i_n} \;
   \mathbf{e}_{i_1=1} \otimes \mathbf{e}_{i_2} \otimes \cdots \otimes \mathbf{e}_{i_n}.
\]

\paragraph{Evolution.}  
The action of an evolution operator on $\mathbf{p}$ is given by
\begin{equation} \label{eq:evolpr}
p_{j_1, \ldots, j_n} \;\;\leftarrow\;\;
\sum_{i_1, \ldots, i_n} 
   b^{\textit{evol}}_{j(j_1, \ldots, j_n),\, i(i_1, \ldots, i_n)} \;
   p_{i(i_1, \ldots, i_n)}
\qquad \forall \{j_1, \ldots, j_n\}.
\end{equation}
This corresponds to a tensor contraction.
We defined
$b^{\mathrm{evol}}_{j,i} = (\mathbf{B}^{\mathrm{evol}})_{j, i}$.

\paragraph{Independence.}  
If the joint distribution factorizes as
\[
p_{i(i_1, i_2, \ldots, i_n)} = \prod_{\ell=1}^n p_{i_\ell},
\]
then the pro-bits are independent, and the tensor $\mathbf{p}$ decomposes into a product of individual pro-bit states.

\subsubsection{Network of Unitary-stochastic Gates}
\label{sec:ndsg}

In a quantum computer, the evolution operator is implemented as a sequence of quantum gates acting on one or a few qubits (Section~\ref{sec:qcc}).  
This idea extends naturally to pro-bits: unitary gates are replaced by \emph{probabilistic gates}, represented by unitary-stochastic matrices.  

For example, consider a gate that acts only on the first two pro-bits while leaving the remaining pro-bits unchanged.  
Its action is specified by
\[
b_{{j(j_1, \ldots, j_n), \, i(i_1, \ldots, i_n)}} = b_{j_1, j_2, i_1, i_2} \; \prod_{\ell=3}^n \delta_{j_\ell, i_\ell},
\]
where $b_{j_1, j_2, i_1, i_2}$ is an entry of a $4 \times 4$ unitary-stochastic matrix, and the Kronecker deltas enforce that untouched pro-bits pass through unchanged.  

If the computation is implemented with $L$ such unitary-stochastic gates, the overall evolution operator can be expressed as
\begin{equation} \label{eq:pcp}
\mathbf{B} = \mathbf{B}^{(L)} \cdots \mathbf{B}^{(2)} \mathbf{B}^{(1)}.
\end{equation}
$\mathbf{B}^{(\ell)} $ described the operation of unitary-stochastic  gate $\ell$.

The resulting network is analogous to a dynamic Bayesian network, with the additional structural constraint that all transition operators are unitary-stochastic.  
This framework also connects naturally to generative models in machine learning, such as diffusion models.  
Furthermore, the transpose operator $\mathbf{B}^\top$ can be realized by transposing each gate and reversing the order, effectively propagating probabilities backward through the network of unitary-stochastic gates.

\subsubsection{Quantum Computer Versus Probabilistic Quantum Computer}

Assume a quantum computer as defined in Equation~\ref{eq:qcp}.  
We can then define
\[
\mathbf{\tilde B} = | \mathbf{U}^{(L)} |.^2 \, \cdots \, | \mathbf{U}^{(2)} |.^2 \, | \mathbf{U}^{(1)} |.^2
\]
which corresponds to a quantum computer in which an ignorant measurement is performed after each quantum gate.   $|\cdot|.^2$ stands for an elementwise magnitude squared operation.
In general, however, $\mathbf{B} \neq \mathbf{\tilde B}$.  
If the goal is to achieve $\mathbf{B} \approx |\mathbf{U}|.^2$, one cannot simply replace quantum gates with unitary-stochastic gates; instead, a redesign of the circuit is required.\footnote{A simple example: Assume that 
\[
\mathbf{U}^{(1)}= \mathbf{U}^{(2)} = \frac{1}{\sqrt{2}}
\begin{bmatrix}
    1       & 1  \\
    1       & -1
\end{bmatrix}
\]
(Hadamard gate). Then $\mathbf{U} = \mathbf{U}^{(2)} \mathbf{U}^{(1)} =   \mathbf{I}$ and this is the same as $|\mathbf{U}|.^2$.
In contrast, $|\mathbf{U}^{(2)}|.^2 |\mathbf{U}^{(1)}|.^2 =
\frac{1}{2}\begin{bmatrix}
    1       & 1  \\
    1       & 1
\end{bmatrix}$. Trivially, $\mathbf{B}^{(1)}= \mathbf{B}^{(2)} = \mathbf{I}$ would be a pro-bit circuit that would result in $|\mathbf{U}|.^2 $.
}

A quantum computer is, of course, realized on physical quantum hardware.  
The same principle applies to a probabilistic quantum computer, although its implementation would be fundamentally different.  
Its key advantage lies in inference: while sampling in a quantum computer is destructive, stochastic sampling in a probabilistic quantum computer is natural, enabling direct inference.  
This property makes probabilistic quantum circuits amenable to efficient realization on classical hardware, for example through neural network architectures.

In the next chapter, we present a construction of a probabilistic quantum computer implemented as a neural network circuit, along with the necessary approximations for its practical realization.

\section{Neural Network Algorithms}
\label{sec:fact}

\subsection{Formal Neurons: Bernoulli Variables and Evolution}
\label{sec:facapprox}
We now show how the TB equations can be derived from the pro-bit representation introduced in Section~\ref{sec_probitr}.  
Each pro-bit can be interpreted as a \emph{formal neuron}.  

If the pro-bits are independent, the joint distribution factorizes as
\begin{equation} \label{eq:sfact}
p_{i(i_1, \ldots, i_n)} 
= \prod_{\ell=1}^n (\gamma_\ell)^{i_\ell} (1-\gamma_\ell)^{1-i_\ell} 
\end{equation}
$$
= \exp\!\Bigg( \sum_{\ell=1}^n 
i_\ell \left[\log(\mathrm{sig}(q_\ell)) -   \log(1-\log(\mathrm{sig}(q_\ell)))\right] +  \log(1-\log(\mathrm{sig}(q_\ell))) \Bigg)
$$
$$
= \exp\!\Bigg( \sum_{\ell=1}^n i_\ell q_\ell - \sum_{\ell=1}^n \log(1+\exp q_\ell) \Bigg)
$$
where the probabilistic state of each pro-bit is specified by a single Bernoulli parameter
\[
\gamma_\ell = \mathrm{sig}(q_\ell), \qquad 0 \le \gamma_\ell \le 1
\]
with $q_\ell$ denoting the logit of $\gamma_\ell$.\footnote{We have used that $q_\ell = \log(\mathrm{sig}(q_\ell)) - \log(1-\mathrm{sig}(q_\ell))$
and that 
$\log(1-\log(\mathrm{sig}(q_\ell))) = 
\log (   1-(1+\exp(-q_{\ell}))^{-1}   ) 
= \log (   (1+\exp q_{\ell})^{-1}  )  = - \log (1 + \exp q_{\ell})$.
}

\subsection{Neural Evolution Operator}
\label{sec:evolnn}

\begin{figure}[h]
 \vspace{+1cm}
\begin{center}
 \includegraphics[width=\linewidth]{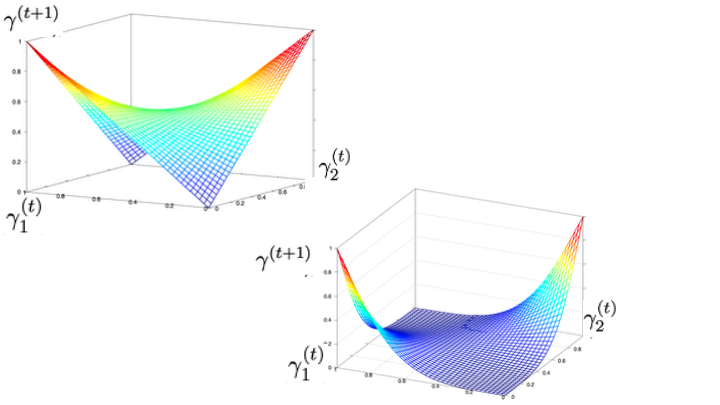}
\end{center}
    \caption{
        Illustrative example with $n=2$. 
        Left: Correct polynomial interpolation. 
        Right: Incorrect interpolation. 
        While the outputs are correct for $\gamma_1, \gamma_2 \in \{0, 1\}$, 
        the interpolation becomes biased for $\gamma_1, \gamma_2 \in (0, 1)$. 
        The right-hand plot demonstrates a form of overfitting, suggesting that regularized solutions are generally preferable. $f^{\textit{NN}^{\textit{evol}} }(\cdot)$ needs to have  large capacity but should be regularized for smooth interpolation. 
    }
\label{fig:xorint}
\end{figure}

We now approximate the evolution operator using four key approximations:

\begin{itemize}

\item \textbf{Approximation 1 — Factorized conditional probability.}  
Starting from Equation~\ref{eq:evolpr}, we impose conditional independence and write
\begin{equation} \label{eq:beintrans2}
b^{\text{evol}}_{j(j_1, \ldots, j_n), \, i(i_1, \ldots, i_n)}
= \prod_{\ell=1}^n 
\beta_{i_1,\ldots,i_n}^{j_\ell}
\big[1-\beta_{i_1,\ldots,i_n}]^{1-j_\ell}
\end{equation}
with Bernoulli parameters $0\le\beta_{i_1,\ldots,i_n}\le 1$.  We only enforce column stochasticity.

\item \textbf{Approximation 2 — Basis-state independence.}   
If the system is initially in a basis state ($\gamma_\ell \in \{0,1\}$), then after evolution the pro-bits remain independent and follow Bernoulli distributions.
In general, contraction introduces dependencies even if the input distribution is  factorized.

\item \textbf{Approximation 3 — Factorized contraction.}  

For each neuron $j$ (after evolution), we obtain a marginal Bernoulli distribution with:
\begin{equation} \label{eq:nnevol3}
\gamma_j \;\leftarrow\;
\sum_{i_1,\ldots,i_n} \left(
\prod_{\ell=1}^n (\gamma_\ell)^{i_\ell}(1-\gamma_\ell)^{1-i_\ell} \right)
\, \beta_{i_1,\ldots,i_n}.
\end{equation}
This is a system of 
$N = 2^n$ basis functions with weights $\beta_{i_1,\ldots,i_n}$. 
The basis functions are  polynomials of degree $n$.

\item \textbf{Approximation 4 — Neural interpolation.}  
 
Instead of enumerating all basis functions, we approximate them using a neural network $f^{\textit{NN}^{\textit{evol}} }(\cdot)$. The neural network is implemented as
\begin{equation} \label{eq:evolout}
\mathbf{h} \;\leftarrow\; \mathrm{sig}(\mathbf{v}_0 + \mathbf{V}\mathbf{q}), \qquad  
\mathbf{q} \;\leftarrow\; \mathbf{W}\mathbf{h}
\end{equation}
where $\mathbf{V}, \mathbf{W}, \mathbf{v}_0$ are trainable weights, $\mathbf{h}$ are hidden activations, and $\vec\gamma = \mathrm{sig}(\mathbf{q})$.  
If the neural network matches Equation~\ref{eq:nnevol3}, this corresponds to a form of Jensen’s approximation\footnote{Jensen’s approximation states that $\sum_x f(x) P(x) \approx f(\sum_x x P(x))$, i.e., the expectation of a function is approximated by the function of the expectation.}, though exact equivalence is not required.  
By the universal approximation theorem, a neural network with at least one hidden layer can approximate the polynomial mapping.  
However, not every interpolating network produces meaningful dynamics (see Figure~\ref{fig:xorint}).

\end{itemize}

\medskip
In summary, the time-evolution equations become \emph{deterministic}:  
we obtain a nonlinear state-space model of dimension $n$, with state vector $\mathbf{q}$.  
The ``noise process’’ appears only when mapping Bernoulli parameters $\gamma_\ell$ to probabilistic states.

\subsection{Neural PVM}
\label{sec:simhbbm}

\subsubsection{Neural PVM}
\label{sec:sconnz}

We consider the one-sided tensorization model introduced in Section~\ref{sec_probitr}.  
A measurement outcome $k$ is represented by a formal neuron in the index layer.  
Upon observing outcome $k$, the post-activation of neuron $k$ is set to $1$, while all other index neurons are set to $0$, except for previously activated indices, which remain at $1$.  

The synaptic weight from index neuron $k$ to representation neuron $i$ is denoted $a_{i,k}$.  
Let $\mathbf{A}$ be the weight matrix with entries $(\mathbf{A})_{i,k} = a_{i,k}$.  
The $k$-th column, $\mathbf{a}_k$, serves as the embedding vector of outcome $k$.  

Conditioned on outcome $k$, we assume a factorized Bernoulli distribution, leading to
\begin{equation} \label{eq:eomh}
\mathbf{q} \;\leftarrow\; \mathbf{a}_k .
\end{equation}
Then, $\vec\gamma = \mathrm{sig}(\mathbf{q})$ and the Bernoulli distribution assumes the form of Equation~\ref{eq:sfact}. Thus, the posterior state is entirely determined by the embedding vector $\mathbf{a}_k$, and the prior probabilistic state is discarded.  
This is analogous to the quantum case: compare Equation~\ref{eq:pkesus} for the PVM posterior and Equation~\ref{eq:pvmpost} for the probabilistic PVM posterior.

\subsubsection{Outcome Probability}

As in the neural evolution operator (Section~\ref{sec:evolnn}), we assume conditional independence.  
Under this assumption, the entries of the $\mathbf{B}$-matrix are given by (recall the identities formulated in Equation~\ref{eq:sfact})
\begin{equation}
\label{eq:bfac}
b_{i(i_1, \ldots, i_n), k} 
= \prod_{\ell=1}^n \big(\mathrm{sig}(a_{\ell,k})\big)^{i_\ell} \big(1 - \mathrm{sig}(a_{\ell,k})\big)^{1-i_\ell} 
\end{equation}
\[= \exp \Bigg( \sum_{\ell=1}^n i_\ell a_{\ell,k} - \log\big(1 + \exp a_{\ell,k}\big) \Bigg).
\]

Suppose now that $\mathbf{B}$ is normalized to be doubly stochastic.  
Such matrices can be obtained either via Sinkhorn's algorithm or as convex combinations of permutation matrices (Birkhoff–von Neumann theorem).  
A simple example of a doubly-stochastic matrix is a permutation matrix.

Using Equation~\ref{eq:propk}, the probability of observing outcome $k$ becomes
\begin{equation} \label{eq:nnwops}
P(k) 
= \sum_{i_1, \ldots, i_n} \exp \Bigg( \sum_{\ell=1}^n i_\ell a_{\ell,k} - \log(1 + \exp a_{\ell,k}) \Bigg) 
\prod_{\ell=1}^n (\gamma_\ell)^{i_\ell} (1-\gamma_\ell)^{1-i_\ell}.
\end{equation}

Applying Jensen's approximation, this can be simplified to
\begin{equation} \label{eq:postpure}
P(k) \approx \exp\Bigg( a_{0,k} + \sum_{\ell=1}^n \gamma_\ell a_{\ell,k} \Bigg),
\end{equation}
where
\[
a_{0,k} = - \sum_{\ell=1}^n \log \big( 1 + \exp a_{\ell,k} \big).
\]

\subsubsection{Neural Postselection}
\label{sec:measbern}

We now consider the calculation of outcome probabilities under postselection.  
Starting from Equation~\ref{eq:povmmea}, the probability of observing the outcome $k$ is
\[
P(k) = \frac{
z_k \sum_{i_1, \ldots, i_n} b_{i(i_1, \ldots, i_n), k} \prod_{\ell=1}^n (\gamma_\ell)^{i_\ell} (1-\gamma_\ell)^{1-i_\ell}
}{
\sum_{k: z_k = 1} \sum_{i_1, \ldots, i_n} b_{i(i_1, \ldots, i_n), k} \prod_{\ell=1}^n (\gamma_\ell)^{i_\ell} (1-\gamma_\ell)^{1-i_\ell}
}.
\]
Applying Jensen's approximation independently to the numerator and denominator gives
\begin{equation} \label{eq:postsk}
P(k) \;\approx\; \frac{
z_k \exp \Big( a_{0,k} + \sum_{\ell=1}^n \gamma_\ell a_{\ell=1,k} \Big)
}{
\sum_{k: z_k = 1} \exp \Big( a_{0,k} + \sum_{\ell=1}^n \gamma_\ell a_{\ell=1,k} \Big)
}
\;\equiv\; z_k \, \mathrm{softmax}^z \Big( a_{0,k} + \sum_{\ell=1}^n \gamma_\ell a_{\ell=1,k} \Big),
\end{equation}
where $\mathrm{softmax}^z$ denotes the softmax restricted to the postselected outcomes.  

Note that the weight parameter from index neuron $k$ to representation neuron $i$ is identical to that from $i$ to $k$, but the transfer functions differ: logistic regression is used for updating the probabilistic state, while the softmax is used for computing outcome probabilities.  
This normalization holds even if $\mathbf{B}$ is only column-stochastic rather than unitary-stochastic.

\subsection{Neural HB-POVM}

\subsubsection{Hadamard Product of Two Bernoulli Distributions}

We make use of the following identity.  
Let $\mathbf{p}_a = \big(\mathrm{sig}(a), \, 1-\mathrm{sig}(a)\big)^\top$ and 
$\mathbf{p}_b = \big(\mathrm{sig}(b), \, 1-\mathrm{sig}(b)\big)^\top$ be two Bernoulli probability vectors.  
Then
\begin{equation} \label{eq:hadsigm}
\frac{\mathbf{p}_a \circ \mathbf{p}_b}{\mathbf{p}_a^\top \mathbf{p}_b} 
= 
\begin{pmatrix}
\mathrm{sig}(a+b) \\[4pt]
1 - \mathrm{sig}(a+b)
\end{pmatrix},
\end{equation}
an identity we refer to as the \emph{Hadamard–logistic rule}.\footnote{
We have used that,  with $c = (1- \exp(-a))^{-1} (1- \exp(-b))^{-1}$, 
$$\frac{\mathrm{sig}(a) \mathrm{sig}(b)}{\mathrm{sig}(a) \mathrm{sig}(b) +
(1-\mathrm{sig}(a)) (1- \mathrm{sig}(b))}  = 
\frac{ c }{
c  + c \exp(-a-b)} = \mathrm{sig}(a+b) .$$
The resulting Bernoulli distribution can also be written as $\gamma_{ab}^i (1-\gamma_{ab})^{1-i}$ with $$\gamma_{ab} = 
 \frac{\gamma_a \gamma_b }{  \gamma_a \gamma_b + (1-\gamma_a)(1- \gamma_b) }$$
and  with $\gamma_a = \mathrm{sig}(a)$ and $\gamma_b = \mathrm{sig}(b)$.
 }

\subsubsection{Neural HB-POVM}
\label{sec:sconnz2}

We now apply this identity to the probabilistic treatment of the HB-POVM.  
The outcome probability $P(k)$ remains unchanged from Equation~\ref{eq:postsk}.

Consider now Equation~\ref{eq:BF1} with the  factorized representations of
Equation~\ref{eq:sfact} and Equation~\ref{eq:bfac}.  When we apply 
the Hadamard–logistic rule (Equation~\ref{eq:hadsigm}) 
we obtain the posterior update
\begin{equation} \label{sec:posteriorskips}
\mathbf{q} \;\leftarrow\; \mathbf{a}_k + \mathbf{q} . 
\end{equation}

Compared with the HB-POVM posterior in Equation~\ref{eq:suppovm} and the probabilistic HB-POVM posterior in Equation~\ref{eq:BF1}, this formulation combines two contributions:  
the prior signal from the probabilistic state and the generative signal from the measurement outcome.  
Structurally, the update mirrors the \emph{skip connection} widely used in ResNets \citep{he2016deep} and transformer architectures, where additive terms preserve and propagate prior information.

Thus, the neural HB-POVM update can be interpreted as a neural PVM update \emph{augmented with a skip connection}.  
This analogy highlights the role of the skip connection as the incorporation of the \emph{logit prior}.

\subsection{Intractable Neural gHMM}
\label{sec:intra}

We first consider the case without postselection.  
If $\mathbf{B}$ is unitary-stochastic and properly normalized, the outcome probability $P(k)$ is obtained directly from Equation~\ref{eq:postpure}, and the probabilistic state is updated according to Equation~\ref{sec:posteriorskips}.  

The case with postselection is treated in Appendix~\ref{sec:intghmm}.  
Using Jensen’s approximation, the outcome probability reduces to Equation~\ref{eq:postsk}, while the posterior distribution becomes
\begin{equation} \label{eq:bnnu}
p_{i(i_1, \ldots, i_n)} \;\leftarrow\;  
  \frac{1}{P(k)} \,
  \mathrm{softmax}^z \!\left( a_{0, k} + \sum_{\ell=1}^n i_\ell a_{\ell=1, k} \right) 
 \prod_{\ell=1}^n (\gamma_\ell)^{i_\ell} (1-\gamma_\ell)^{1-i_\ell}.
\end{equation}
The product is the prior and the softmax the likelihood. A detailed derivation of this expression is provided in Appendix~\ref{sec:intghmm}.

In contrast to the neural HB-POVM, where pro-bits remain independent, the posterior in this case generally exhibits dependencies across pro-bits.  
Consequently, while the neural HB-POVM with postselection remains analytically tractable, Bayesian updating in a generalized HMM (gHMM) with postselection quickly becomes intractable.  
To address this difficulty, one typically resorts to variational methods or related approximation techniques to obtain practical solutions.  

\medskip

\begin{table}[ht]
\centering

\caption{
Overview of algorithms.   The first column presents the \textit{PVM family} (quantum, probabilistic, neural) and 
the second column presents the \textit{POVM family} (quantum, probabilistic, neural). 
As comparison: The third column presents the gHMM. The fourth column describes  the generative recurrent neural network: 
there is no feed back from the measurement outcome to the state and $Y$-measurements are independent. (ps.) stands for postselection. The TB uses HB-POVM, rows 8 and 9 (blue); the intractable Bayes update is gHMM line 9 (red).
}
\begin{tabular}[t]{| c | l | l l c  c |}

\hline

 & & PVM  & HB-POVM &  gHMM  & gRNN\\
 \hline
  \hline
  
 & \textbf{Quantum:}  &&& &\\
1& $P(k)$  					& Eq.~\ref{eq:pkele} 		& Eq.~\ref{eq:pkhbo} & --    & -- \\
2 & $P(k)$ with posts. 				& Eq.~\ref{eq:pspvm}	& Eq.~\ref{eq:povmmea}& --  & -- \\
3 & $\mathbf{\vec \psi}$ update 	& Eq.~\ref{eq:pkesus} 	& Eq.~\ref{eq:suppovm} &  -- & -- \\

\hline
& \textbf{Probab.:}  &&&& \\
4 & $P(k)$  			& Eq.~\ref{eq:propk} 		& Eq.~\ref{eq:propk}		
& Eq.~\ref{eq:propk} & -- \\
5 & $P(k)$ with posts.  		& Eq.~\ref{eq:povmmea}	& Eq.~\ref{eq:povmmea} 	
& Eq.~\ref{eq:bmeass} & -- \\
6 & $\mathbf{p}$ update 	& Eq.~\ref{eq:pvmpost} 	& Eq.~\ref{eq:BF1} 		
&  Eq.~\ref{eq:BF1}/\ref{eq:bmeass987}(ps.) & -- \\

\hline

&  \textbf{Neural:}  &&& & \\
 7 & $P(k)$  			& Eq.~\ref{eq:postpure} 	& Eq.~\ref{eq:postpure}					
 & Eq.~\ref{eq:postpure}				& Eq.~\ref{eq:argmax} \\
8 &  $P(k)$ with posts. 		& Eq.~\ref{eq:postsk} 	& {\color{blue}  Eq.~\ref{eq:argmax}}			 
& Eq.~\ref{eq:postsk} 			& Eq.~\ref{eq:argmax} \\
 
9 & $\mathbf{q}$ update 	& Eq.~\ref{eq:eomh}		&{\color{blue} Eq.~\ref{sec:posteriorskips}} 	
 &Eq.~\ref{sec:posteriorskips} / {\color{red} \ref{eq:bnnu} (ps.)}	& $\vec \gamma \leftarrow \vec \gamma$\\

\hline
\end{tabular}
\label{tab:overv}
\end{table}%

\section{Towards the TB Algorithm}
\label{sec:back}

Table~\ref{tab:overv} summarizes the main results, which we now place in the broader context of the TB algorithm introduced in Section~\ref{sec:tbt}.

\subsection{Generative Measurement}

Algorithm~\ref{alg:cap-3NN} realizes generative measurement through a neural HB-POVM with postselection.  
Concretely, measurement is performed according to Equation~\ref{eq:postsk}, followed by the state update in Equation~\ref{sec:posteriorskips}.

\subsection{External Inputs} 
\label{sec:exti2}

We now extend the framework to incorporate input as  $\vec \nu$.  $\vec \nu$ denotes low-level cortical areas such as  primary sensory cortices,  early sensory regions, and maybe even subcortical regions. In the further discussion we typically assume that these are brain states driven by visual or other sensory inputs, such as V1 and V2.
Formally, we update the state as
\begin{equation} \label{eq:eom333}
\mathbf{q} \;\leftarrow\; \mathbf{q} + \mathbf{g}(\vec \nu).
\end{equation}
Thus, external inputs enter the dynamics through an additional skip connection, as implemented in Algorithm~\ref{alg:cap-2NN}.  

From a probabilistic perspective, this construction is equivalent to treating an external input as the outcome of an HB-POVM measurement event, with embedding vector
\[
\mathbf{g}(\vec \nu) = \bigl(g_{1}(\vec \nu), \dots, g_{n}(\vec \nu)\bigr)^\top.
\]
In this way, external signals are naturally fused with the internal state update, ensuring seamless integration of input-driven and intrinsic dynamics.
$\vec \nu$ is not restricted to sensory inputs but could represent any internal brain state.

The (approximate) mapping from index $k$ to 
 $\mathbf{q} = \mathbf{a}_k$  and back to $\vec \nu_k$ is an embodiment process. 
It   could be realized by a network implementing a function  where
$\hat {\vec \nu}_k \leftarrow \mathbf{g}^+(\mathrm{sig}(\mathbf{a}_k))$ is the embodiment of the index $k$.  The map $\vec \nu_k \mapsto \mathbf{a}_k 
\mapsto \hat {\vec \nu}_k$ forms an autoencoder structure.

\subsection{Modes of Operations}
\label{sec:measneurins}

We consider a generative outcome $k$. The pre-CBS update takes the form
\[
\mathbf{q} \;\leftarrow\;   
\alpha \bigl(\mathbf{W} \mathbf{h} + \mu \mathbf{g}(\vec \nu) \bigr) 
+ \beta \,\mathbf{a}_k .
\]
Three parameters govern the dynamics.  
In the TB model (neural HB-POVM), we set $\alpha=1$, so the state is preserved, and $\beta=1$, so the outcome feeds back to the pre-CBS.  
The parameter $\mu$ controls the influence of sensory input.  

When several symbolic labels are generated with $\alpha=\beta=1$, we obtain
\[
\mathbf{q} \;\leftarrow\; \mathbf{W}\mathbf{h} + \mathbf{g}(\vec \nu) + \sum_{k \in S} \mathbf{a}_k ,
\]
where $S$ denotes the set of active labels (e.g., for a region of interest in a scene).  
This expression shows that the firing rates from the hidden state $\mathbf{h}$, the sampled outcomes $k \in S$, and the sensory input remain stable between two evolution steps.  

\paragraph{Perception.}  
In perception, all three parameters are set to $\alpha=\beta=\mu=1$.  
Here, perceptual input drives the pre-CBS, and the TB dynamics are oriented toward the environment.  
The sequence of operations is:  
\begin{itemize}
\item  Evolution via Algorithm~\ref{alg:cap-1NN},  
\item Input absorption and attention via Algorithm~\ref{alg:cap-2NN}, where sensory signals (e.g., visual input from a region of interest) are added,   
\item   Generative measurement via Algorithm~\ref{alg:cap-3NN}, which identifies the concept of interest in the context of past inputs.  
\end{itemize}

Algorithm~\ref{alg:cap-3NN} can be iterated to identify related concepts within the same region of interest (ROI).  
Subsequent TB evolution shifts the focus to the next ROI.  
For example, in visual perception the first ROI is the whole scene (yielding scene labels), the second ROI focuses on a subject $s$, the third on an object $o$, and the fourth on the predicate $p$.  With $\beta = 0$, the brain can completely focus on perceptual inputs. 

\paragraph{Memory.}  
During memory recall, $\mu=0$ and sensory input is ignored.  
Intermediate modes are also possible, where memory dominates but external input retains partial influence.  

The TB can also operate in a neural PVM mode by setting $\alpha=0$, $\beta=1$.  
In this case, the update reduces to $\mathbf{q} \leftarrow \mathbf{a}_k$, i.e., the pre-CBS focuses on a recalled concept while ignoring perceptual context.  

\paragraph{Generative Recurrent Neural Network (gRNN).}  
Finally, we consider a generative recurrent neural network (gRNN), obtained by setting $\beta=0$, so that measurement outcomes do not feed back into the state.  
In this case, one may avoid sampling and instead deterministically select
\begin{equation} \label{eq:argmax}
 k = \arg \max_{k'} P(k'),   
\end{equation}
yielding a deterministic gRNN.

\subsection{Ignorant $Y$-measurement   and Attention}
\label{sec:igg}

We now consider the case of an \emph{ignorant $Y$-measurement}.  
The corresponding update rule is
\begin{equation} \label{eq:igg}
\mathbf{q} \;\leftarrow\; \mathbf{q}  + 
\sum_{k: z_k =1} \mathbf{a}_{k} \,
\mathrm{softmax}^z\!\left(a_{0, k} + 
\sum_{\ell=1}^n \gamma_\ell a_{\ell=1, k}\right),
\end{equation}
which directly realizes the attention mechanism of Algorithm~\ref{alg:cap-2NN}.  
The derivation of this equation can be found in Appendix \ref{sec:igna}.

Attention can be directed toward past episodic memories, semantic concepts, or symbolic labels.  
Analogous to machine learning architectures, attention stabilizes predictions and resolves contextual ambiguity.

\subsection{Evolution Operator}

The evolution operator is implemented as described in Algorithm~\ref{alg:cap-1NN}.  
A natural refinement of the neural network evolution operator is to introduce ResNet-style skip connections by \textit{augmenting the $\mathbf{h}$-layer}.  
Such connections improve stability and expressivity while preserving the overall architecture of the operator.

\subsection{The Dynamic Context Layer, Working Memory and the Global Workspace, Revisited}



We propose that the evolution operator $f^{\textit{NN}^{\textit{evol}}}(\cdot)$ represents a relatively recent addition to the brain’s architecture in biological evolution. This operator interrupts the integration phase, enabling the system to shift attention to a new concept or maintain focus on the current one. It functions by compressing the Cognitive Brain State (CBS) and resetting all symbolic indices to an inactive state.

We further suggest that the dynamic context layer—the $\mathbf{h}$-layer—has a dimensionality much smaller than that of the representation layer ($n_H \ll n$), making it a genuine bottleneck. Consistent with the interpretation of the representation layer as the global workspace—the brain’s representational canvas—we assume that a cognitive experience engages large, distributed regions of the brain and remains grounded in perception and action. \cite{baars1997theater} and \cite{dehaene2014consciousness} have argued that the global workspace itself constitutes a bottleneck; accordingly, the dynamic context layer may serve as another candidate for this functional role.

The dynamic context layer may also underlie working memory. We further propose that the brain can switch among different evolution operators, $f^{\textit{NN}^{\textit{evol}}}(\cdot)$, under cognitive control. Some of these operators may comprise multiple deep neural layers, thereby supporting higher levels of intelligent processing.

\section{Discussion}
\label{sec:discuss}

\subsection{Computation and the Brain} 

\subsubsection{Quantum Computing and the Brain}

From the perspective of this work, quantum-state computation may confer distinct advantages for certain classes of problems.  
Whether the brain in fact exploits such quantum computational mechanisms remains an open question, and skepticism regarding this possibility is widespread among researchers \citep{tegmark2000importance}.

\subsubsection{Probabilistic Quantum Computing and the Brain}

In an \textit{epistemic} interpretation of the probabilistic quantum equations, the brain would still be required to implement the underlying quantum dynamics—operating with genuine quantum states and unitary gates.  
This view implies that the brain effectively functions as a quantum computer, as discussed in the previous section.  

In contrast, under an \textit{ontic} interpretation, the brain need only realize the corresponding probabilistic quantum algorithms, using probabilistic states and unitary-stochastic gates.  
This latter interpretation is the one we consider more plausible.  


\subsubsection{Neural Quantum Computing and the Brain}

We propose that the TB algorithms may more closely approximate neural operations in the brain than either quantum or probabilistic quantum models.  
In this framework, the CBS vector $\vec{\gamma}$ can be interpreted as the firing rates of neurons in the representation layer (rate coding).  
A similar rate-coding scheme applies to the neurons in the hidden layer $\mathbf{h}$ of the evolution network.  
Meanwhile, the selected binary index neurons remain active between successive neural evolution steps, maintaining continuity of symbolic information.  


\subsubsection{Recurrent Neural Networks and the Brain}

A recurrent neural network (gRNN) may suffice for perceptual processing but cannot realize memory functions, as it lacks connections from the index neurons to the representation layer.  
In the non-probabilistic (deterministic) limit, the gRNN does not sample outcomes; instead, they are deterministically specified by the current CBS state (compare Equation~\ref{eq:argmax}).

\subsection{The TB and Transformer LLMs}  
\label{sec:llmtb}

There are numerous structural and functional parallels between the Tensor Brain (TB) framework and transformer-based large language models (LLMs).  
In both systems, a symbol in the TB corresponds to a token in an LLM.  
Each employs neural networks, skip connections, and attention mechanisms, although their specific implementations differ in detail.  

A key distinction is that, in the TB, the representation layer and the cognitive basis state (CBS) play a central role in information processing.  
In transformer architectures, a comparable role may be played by the final hidden layer, which encodes high-level semantic representations.  

Memory integration in the TB is conceptually analogous to retrieval-augmented generation (RAG) in LLMs: semantic memory provides conceptual knowledge, while episodic memory contributes context from past experiences.  
The crucial difference is that RAG retrieves and incorporates explicit text data, whereas the TB dynamically generates symbolic indices from learned memory embedding vectors.

\subsection{Explicit and Distributed Indices}
\label{sec:distrind}

\subsubsection{Explicit Indices}

In the TB framework, each symbol is represented by a dedicated index neuron.  
Introducing a new symbol therefore entails recruiting an additional index neuron.  
Initially, the system may contain a large pool of up to $N$ index neurons with randomly initialized embeddings, but only the postselected outcomes $k: z_k = 1$ correspond to known symbols. 
When a new symbol is encountered, an unused index neuron from the random pool can be recruited and assigned an embedding vector, thereby extending the symbolic repertoire dynamically.

\subsubsection{Distributed Indices (Two-sided Tensorization)}

In distributed representations, explicit index neurons are replaced by a population of $n$ binary index neurons, which encode information through population coding (see the discussion on two-sided tensorization in Section~\ref{sec:one-sided} and the pro-bit representation in Section~\ref{sec_probitr}).  
In the probabilistic PVM (Figure~\ref{fig:PVM}, top), the mappings between the distributed $X$ and $Y$ variables can be implemented by networks of unitary-stochastic gates, allowing straightforward inference via sampling.  

Alternatively, we may define the embedding vectors as  
\[
\mathbf{a}_k \;\gets\; \mathbf{f}^{\textit{NN}}(y_{1, k}, \ldots, y_{n, k}),
\]
where $y_{1, k}, \ldots, y_{n, k}$ with $y_{\ell} \in \{0, 1\}$ represent the activations of the  index neurons, and  
$\mathbf{f}^{\textit{NN}}(\cdot)$ denotes a neural network mapping  
$\{0, 1\}^n \rightarrow \mathbb{R}^n$.  
This substitution applies to all equations involving $\mathbf{a}_k$, particularly Equations~\ref{eq:postsk} and~\ref{sec:posteriorskips}.

To store, address, or retrieve a distributed index pattern, one can allocate an additional set of explicit index neurons.
Associative memory models such 
modern Hopfield networks~\citep{ramsauer2020hopfield} need explicit indices as well; there, they are called memory neurons. In the TB, the iterative update in modern Hopfield networks is replaced by the sampling of the explicit indices. 

In Section \ref{sec:exti2}, we have argued that in perception
$\mathbf{q} \leftarrow \mathbf{g}(\vec \nu)$ is a map from primary sensory cortices to the representation layer and that $\hat {\vec \nu} = \mathbf{g}^+(\mathrm{sig}(\mathbf{q}))$ is the approximate reverse mapping.  It is naturally to think of $\vec \nu_k$ as a distributed index for $\mathbf{a}_k$. Thus, 
maybe distributed indices do not require separate neural resources and $\mathbf{y} \equiv \vec \nu$.

\subsection{Quantum Decoherence and Self-Measurements}

A central issue addressed in this paper is the modeling of the measurement process.  
The measurement postulate (Postulate~3) in quantum mechanics remains the subject of continuing debate.  
As noted by \citet[p.~85f]{nielsen2010quantum}:
\begin{quote}
``Measuring devices are quantum mechanical systems, so the quantum system being measured and the measuring device together are part of a larger, isolated, quantum mechanical system. Might it be possible to derive Postulate~3 as a consequence of this picture? Despite considerable investigation along these lines, there is still disagreement between physicists about whether or not this is possible.''
\end{quote}

Decoherent quantum theory (see Section~\ref{sec:qdec}) approaches this problem by modeling the system of interest as entangled with a much larger environment.  
Within this framework, observations of a subsystem appear probabilistic, even though the joint system evolves unitarily.  
Some researchers argue that nonselective measurement provides the operative mechanism behind decoherence: information about specific outcomes becomes entangled with environmental degrees of freedom and is effectively irretrievable.  

From this perspective, the brain itself may function as such an environment—coupling to smaller subsystems and thereby acting as the agent that initiates measurement-like processes.  
In this view, the probabilistic quantum updates in the TB could arise naturally from the brain’s internal dynamics and its interaction with its own environment, without the need to invoke an external observer.

\subsection{Relationship to Quantum Cognition}

Quantum cognition explores whether the mathematical formalism of quantum theory can account for non-Bayesian aspects of human reasoning—such as order effects, contextuality, and interference phenomena.

\subsubsection{Order Effect}

\paragraph{PVM.}

Quantum projection operators are idempotent: repeating the same measurement yields the same outcome, leaving the quantum state unchanged after the first measurement.  
However, sequential measurements in different, non-commuting bases exhibit \textit{order effects}.  
Let $k$ and $k'$ denote outcomes measured in bases $\mathbf{U}$ and $\mathbf{U}'$, respectively.  
For a quantum state $\vec{\psi}$, we have
\[
P(k) = |\mathbf{u}_k^{\dagger} \vec{\psi}|^2
\quad \text{and} \quad
P(k') = |(\mathbf{u}'_{k'})^{\dagger} \vec{\psi}|^2.
\]
Then,
\[
P(k \text{ then } k') = P(k)\; |(\mathbf{u}'_{k'})^{\dagger} \mathbf{u}_{k}|^2,
\qquad \vec{\psi} \leftarrow \mathbf{u}'_{k'},
\]
is generally \textit{not} the same as
\[
P(k' \text{ then } k) = P(k')\; |\mathbf{u}_{k}^{\dagger} \mathbf{u}'_{k'}|^2,
\qquad \vec{\psi} \leftarrow \mathbf{u}_{k}.
\]
We thus observe a clear order effect: the resulting posterior quantum states differ depending on the measurement sequence.

This phenomenon carries over to the {probabilistic PVM}.  
With $P(k) = \mathbf{b}_k^{\top} \mathbf{p}$ and $P(k') = (\mathbf{b}'_{k'})^{\top} \mathbf{p}$, we obtain
\[
P(k \text{ then } k') = P(k)\, (\mathbf{b}'_{k'})^{\top} \mathbf{b}_{k},
\qquad \mathbf{p} \leftarrow \mathbf{b}'_{k'},
\]
and
\[
P(k' \text{ then } k) = P(k')\, \mathbf{b}_{k}^{\top} \mathbf{b}'_{k'},
\qquad \mathbf{p} \leftarrow \mathbf{b}_{k}.
\]
Again, the order of measurements leads to different outcomes, confirming the persistence of order effects.  

\paragraph{HB-POVM.}

For the HB-POVM, we obtain
\[
P(k, k') = (\mathbf{b}'_{k'})^\top (\mathbf{b}_k \circ \mathbf{p}),
\qquad
\vec{\psi} \leftarrow
\frac{\mathbf{u}'_{k'} \circ \mathbf{u}_{k} \circ \vec{\psi}}
{\sqrt{(\mathbf{b}'_{k'})^\top (\mathbf{b}_k \circ \mathbf{p})}}.
\]
This expression is \textit{independent} of the order of measurements.  

Similarly, for the probabilistic HB-POVM,
\[
P(k, k') = (\mathbf{b}'_{k'})^\top (\mathbf{b}_k \circ \mathbf{p}),
\qquad
\mathbf{p} \leftarrow
\frac{\mathbf{b}'_{k'} \circ \mathbf{b}_k \circ \mathbf{p}}
 {(\mathbf{b}'_{k'})^\top (\mathbf{b}_k \circ \mathbf{p})}\]
which is equivalent to a Bayes update in a generalized HMM (gHMM).

\paragraph{Neural Network.}

For the {neural PVM}, the posterior takes the form  
$\mathbf{q} \leftarrow \mathbf{a}_k$ if $k$ follows $k'$,  
and $\mathbf{q} \leftarrow \mathbf{a}_{k'}$ if $k'$ follows $k$—clearly exhibiting an order effect.  
In contrast, for the {neural HB-POVM}, the update  
$\mathbf{q} \leftarrow \mathbf{a}_{k'} + \mathbf{a}_{k''} + \mathbf{q}$  
shows \textit{no} order dependence.

Table~\ref{tab:pvm_povm} and Table~\ref{tab:roi_comparison} in Appendix~\ref{sec:expresul} provide experimental results on the order effect.

\paragraph{Summary.}

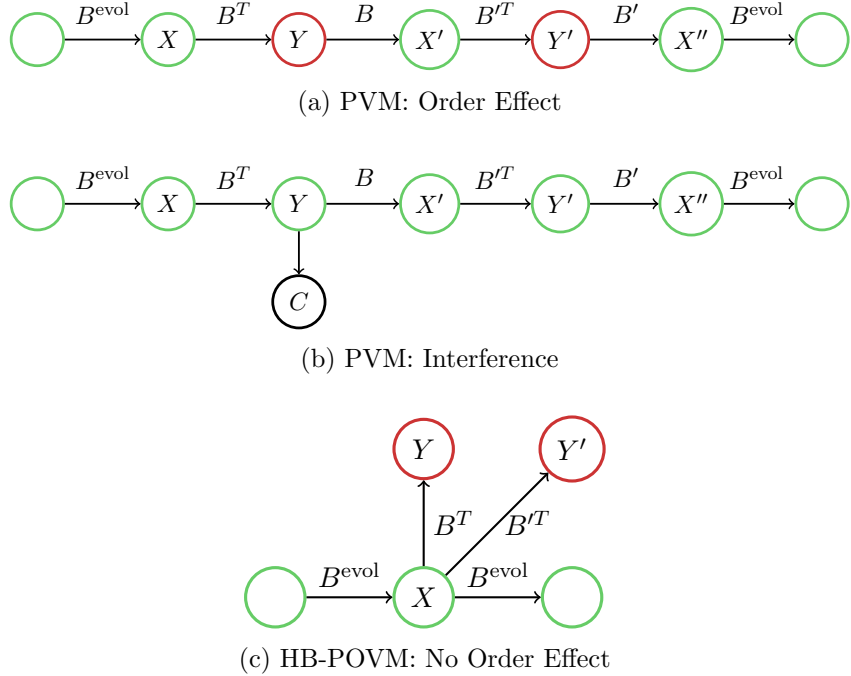
\begin{figure}[hbt!]
    \centering
    \begin{subfigure}[b]{0.9\textwidth}
        \centering
        \resizebox{\linewidth}{!}{
        \begin{tikzpicture}[
        xstate/.style={circle, draw=lightgreen, very thick, minimum size=0.8cm, fill=white},
        ystate/.style={circle, draw=darkred, very thick, minimum size=0.8cm, fill=white},
        arrow/.style={->, thick, darkblue},
        dotted_arrow/.style={->, thick, dotted, darkblue},
        evolution_arrow/.style={->, thick, black},
        label/.style={font=\large\bfseries},
        sublabel/.style={font=\normalsize},
        postselect/.style={circle, draw=darkblue, thick, minimum size=0.8cm, fill=lightblue!20, inner sep =1pt}]

        \node[xstate] (X-1) at   (-6, 0) {};
        \node[xstate] (X) at     (-4, 0) {$X$};
        \node[ystate] (Y) at     (-2, 0) {$Y$};
        \node[xstate] (X') at    (0, 0)  {$X'$};
        \node[ystate] (Y') at    (2, 0)  {$Y'$};
        \node[xstate] (X'') at   (4, 0)  {$X''$};
        \node[xstate] (X''+1) at (6, 0)  {};
        
        \draw[evolution_arrow] (X-1) -- (X);
        \draw[evolution_arrow] (X) -- (Y);
        \draw[evolution_arrow] (Y) -- (X');
        \draw[evolution_arrow] (X') -- (Y');
        \draw[evolution_arrow] (Y') -- (X'');
        \draw[evolution_arrow] (X'') -- (X''+1);
        
        \node[sublabel] at (-5, 0.4) {$B^{\text{evol}}$};
        \node[sublabel] at (-3, 0.4) {$B^T$};
        \node[sublabel] at (-1, 0.4) {$B$};
        \node[sublabel] at (1, 0.4) {$B'^T$};
        \node[sublabel] at (3, 0.4) {$B'$};
        \node[sublabel] at (5, 0.4) {$B^{\text{evol}}$};
        
        \end{tikzpicture}
        }
        \caption{PVM: Order Effect}
        \label{fig-prob-pvm-2F-order}
    \end{subfigure}

    \hspace{2pt}

        \begin{subfigure}[b]{0.9\textwidth}
        \centering
        \resizebox{\linewidth}{!}{
        \begin{tikzpicture}[
        cstate/.style={circle, draw=black, very thick, minimum size=0.8cm, fill=white},
        xstate/.style={circle, draw=lightgreen, very thick, minimum size=0.8cm, fill=white},
        ystate/.style={circle, draw=darkred, very thick, minimum size=0.8cm, fill=white},
        arrow/.style={->, thick, darkblue},
        dotted_arrow/.style={->, thick, dotted, darkblue},
        evolution_arrow/.style={->, thick, black},
        label/.style={font=\large\bfseries},
        sublabel/.style={font=\normalsize},
        postselect/.style={circle, draw=darkblue, thick, minimum size=0.8cm, fill=lightblue!20, inner sep =1pt}]

        \node[xstate] (X-1) at   (-6, 0) {};
        \node[xstate] (X) at     (-4, 0) {$X$};
        \node[xstate] (Y) at     (-2, 0) {$Y$};
        \node[xstate] (X') at    (0, 0)  {$X'$};
        \node[xstate] (Y') at    (2, 0)  {$Y'$};
        \node[xstate] (X'') at   (4, 0)  {$X''$};
        \node[xstate] (X''+1) at (6, 0)  {};
        
        \draw[evolution_arrow] (X-1) -- (X);
        \draw[evolution_arrow] (X) -- (Y);
        \draw[evolution_arrow] (Y) -- (X');
        \draw[evolution_arrow] (X') -- (Y');
        \draw[evolution_arrow] (Y') -- (X'');
        \draw[evolution_arrow] (X'') -- (X''+1);
        
        \node[sublabel] at (-5, 0.4) {$B^{\text{evol}}$};
        \node[sublabel] at (-3, 0.4) {$B^T$};
        \node[sublabel] at (-1, 0.4) {$B$};
        \node[sublabel] at (1, 0.4) {$B'^T$};
        \node[sublabel] at (3, 0.4) {$B'$};
        \node[sublabel] at (5, 0.4) {$B^{\text{evol}}$};

        \node[cstate] (C) at (-2, -1.5) {$C$};
        \draw[evolution_arrow] (Y) -- (C);
        
        \end{tikzpicture}
        }
        \caption{PVM: Interference}
        \label{fig-prob-pvm-2F}
    \end{subfigure}

    \hspace{2pt}
    
    \begin{subfigure}[b]{0.4\textwidth}
        \centering
        \resizebox{\linewidth}{!}{
        \begin{tikzpicture}[
        xstate/.style={circle, draw=lightgreen, very thick, minimum size=0.8cm, fill=white},
        ystate/.style={circle, draw=darkred, very thick, minimum size=0.8cm, fill=white},
        arrow/.style={->, thick, darkblue},
        dotted_arrow/.style={->, thick, dotted, darkblue},
        evolution_arrow/.style={->, thick, black},
        label/.style={font=\large\bfseries},
        sublabel/.style={font=\normalsize},
        postselect/.style={circle, draw=darkblue, thick, minimum size=0.8cm, fill=lightblue!20, inner sep=1pt}
        ]
        \node[xstate] (X-1) at (-2, 0) {};
        \node[xstate] (X) at (0, 0) {$X$};
        \node[xstate] (X+1) at (2, 0) {};
        
        \draw[evolution_arrow] (X-1) -- (X);
        \draw[evolution_arrow] (X) -- (X+1);
        
        \node[sublabel] at (-1, 0.3) {$B^{\text{evol}}$};
        \node[sublabel] at (1, 0.3) {$B^{\text{evol}}$};
        
        \node[ystate] (Y) at (0, 2) {$Y$};
        \draw[evolution_arrow] (X) -- (Y);
        \node[sublabel] at (0.4, 1) {$B^T$};
        \node[ystate] (Y') at (2, 2) {$Y'$};
        \draw[evolution_arrow] (X) -- (Y');
        \node[sublabel] at (1.4, 1) {$B'^T$};
        
        \end{tikzpicture}
        }
        \caption{HB-POVM: No Order Effect}
        \label{fig:y-meas-hb-povm}
    \end{subfigure}
    \hfill
    
    \caption[foo bar]{
{Order effects and interference in probabilistic quantum models.}   
\textit{Top row:} The PVM—both quantum and probabilistic—exhibits order effects, since exchanging $Y$ and $Y'$ yields different outcomes.  
In the quantum case, this corresponds to measurements in different bases.  
\textit{Middle row:} The quantum PVM displays interference, whereas the probabilistic PVM does not.  
Here, the cluster variable $C$ is modeled as a child of $Y$.  
\textit{Bottom row:} The HB-POVM shows no order effect, and the probabilistic HB-POVM exhibits no interference.  $\textbf{B}'$ might be different from $\textbf{B}$.
}
\label{fig:PVM2}
\end{figure}

In summary, order effects arise in quantum, probabilistic, and neural PVMs, but are absent in their HB-POVM counterparts. The underlying dependency structures are depicted in Figure~\ref{fig:PVM2}.

\subsubsection{Order Effects in Relationships  between Concepts}

In all TB models, the system focuses on a single concept between two consecutive evolution operations.  
Relations between concepts are then analyzed through the evolution operator, which mediates their interactions.  
This framework naturally introduces asymmetries in reasoning---for example, the probability of \textit{John likes Mary} may differ from that of \textit{Mary likes John}.  
However, symmetric relations, such as \textit{isMarriedTo}, can be learned through appropriate parameterization of the evolution neural network.

\subsubsection{Experimental Results on Order Effects}
\label{sec:expresul-haupt}

Our experimental results (Table~\ref{tab:pvm_povm-haupt}) confirm that the PVM model exhibits order effects, whereas the HB-POVM is relatively insensitive to the order. A detailed description can be found in Appendix~\ref{sec:expresul}. 


\begin{table*}[!h]
\centering
\caption{Comparison of order sensitivity between PVM and HB-POVM on 10k ImageNet samples. Divergence metrics (KL, JSD) and label reversal rates reveal that PVM is strongly affected by query order, whereas HB-POVM remains largely invariant.}
\label{tab:pvm_povm-haupt}
\renewcommand{\arraystretch}{1.2}
\setlength{\tabcolsep}{5pt}
\begin{tabular}{lcc}
\toprule
\textbf{Metric} & \textbf{PVM} & \textbf{HB-POVM} \\
\midrule
KL & 20.54 & 0.304 \\
JSD & 0.572 & 0.037 \\
\bottomrule
\end{tabular}
\end{table*}

\subsubsection{Quantum Interference}
\label{sec:qinf}

In quantum cognition, quantum interference effects are often linked to the \emph{disjunction fallacy} and to violations of the law of total probability.  
We now analyze the disjunction fallacy within the framework of our models.

\paragraph{PVM.}  
Let $P(k') = |\mathbf{u}_{k'}^{\dagger} \vec{\psi}|^2$,  
$P(k'') = |\mathbf{u}_{k''}^{\dagger} \vec{\psi}|^2$,  
$P(k|k') = |(\mathbf{u}_{k})^{\dagger} \mathbf{u}_{k'}|^2$, and  
$P(k|k'') = |(\mathbf{u}_{k})^{\dagger} \mathbf{u}_{k''}|^2$.  
Then the disjunction probability becomes
\[
P(k \mid k' \text{ OR } k'') 
= 
\frac{
P(k') P(k|k') + P(k'') P(k|k'') + \mathrm{Interference}
}{
P(k') + P(k'')
}.
\]
The quantum interference term is (see Appendix~\ref{sec:intf})
$$\mathrm{Interference} = 2 \mbox{ Re } 
\left(
(\mathbf{u}_{k'}^{\dagger}  \vec \psi) (\mathbf{u}_{k}^{\dagger}   \mathbf{u}_{k'})  
( (\mathbf{u}_{k''}^{\dagger}  \vec \psi )(   \mathbf{u}_{k}^{\dagger}  \mathbf{u}_{k''} ) )^*
\right) .$$

\paragraph{Probabilistic PVM.}
For the probabilistic PVM, we have  
$P(k') = \mathbf{b}_{k'}^{\top} \mathbf{p}$,  
$P(k'') = \mathbf{b}_{k''}^{\top} \mathbf{p}$,  
$P(k|k') = (\mathbf{b}_{k})^{\top} \mathbf{b}_{k'}$, and  
$P(k|k'') = (\mathbf{b}_{k})^{\top} \mathbf{b}_{k''}$.  
The disjunction probability is then purely classical:
\[
P(k \mid k' \text{ OR } k'') = 
\frac{
P(k') P(k|k') + P(k'') P(k|k'')
}{
P(k') + P(k'')
}.
\]

\paragraph{Discussion.}
We clearly observe quantum interference in the quantum PVM model but not in the probabilistic PVM.  
If we regard the latter as implementing correct Bayesian inference, then the quantum PVM indeed exhibits the \emph{disjunction fallacy}.  

The quantum interference term in the quantum PVM apparently  violates the assumption that measurement outcomes can be treated probabilistically.  
We propose that disjunctions should instead be modeled by introducing an additional categorical variable $C$, as illustrated in Figure~\ref{fig:PVM2}.  This leads to a different model (``a different protocol'') which then, again, can be treated probabilistically. See Appendix~\ref{sec:dijfal}.

\paragraph{Probabilistic HB-POVM and Bayes Update in the gHMM.}
For the probabilistic HB-POVM and for the Bayesian update in the generalized HMM (gHMM), the posterior takes the same form:
\[
P(k \mid k' \text{ OR } k'') =
\frac{
\mathbf{p}^{\top} \!\bigl( \mathbf{b}_{k} \circ (\mathbf{b}_{k'} + \mathbf{b}_{k''}) \bigr)
}{
P(k') + P(k'')
}.
\]
Again, there is no interference effect in this case.

\section{Conclusion}
\label{sec:concl}

We began by reviewing quantum systems and then showed how \emph{probabilistic quantum systems} can be derived, which admit efficient implementations using neural network architectures.  
We further argued that such neural implementations may correspond to the TB model—a framework for describing perception and memory in the brain.

We have discussed quantum inference, probabilistic quantum inference, and Bayesian reasoning in canonical generalized hidden Markov models (gHMMs), and compared the PVM and HB-POVM formulations.  
Quantum PVMs exhibit the hallmark features of quantum systems, including order effects and interference.  
Probabilistic PVMs retain order effects but eliminate interference, aligning them more closely with classical probabilistic reasoning.

In probabilistic quantum models, the $\mathbf{B}$ matrix is unitary-stochastic, and inference in a probabilistic HB-POVM is formally equivalent to Bayesian reasoning in a gHMM.  
In more general cases, however—particularly under postselection—the two frameworks diverge.  
While probabilistic HB-POVM reasoning remains computationally tractable, exact Bayesian updates in gHMMs quickly become intractable.

Our analysis suggests that the probabilistic HB-POVM offers distinct advantages in both computational efficiency and biological plausibility.  
It provides a tractable, interpretable framework for reasoning about dependencies between measurement outcomes, making it a strong candidate for modeling information processing mechanisms in the brain.

\bibliography{QuantumTensorBrainSem}

\section*{APPENDIX}

\section{Appendix: Ignorant $Y$-measurement and Attention}
\label{sec:igna}

Here, we derive Equation \ref{eq:igg} in Section \ref{sec:igg}.

\subsection{Quantum}

In quantum, a non-measurement has no effect. This is different for an ignorant measurement. We already discussed ignorant measurements in the $X$-basis in Section \ref{sec:uuoo}. 
Here, we discuss ignorant measurements in the $Y$-basis. 
Let's assume that the system initially  is in quantum state $\vec \psi$. 

\paragraph{Quantum PVM} For an ignorant measurement in the $Y$-basis, we get, 
using Equation \ref{eq:ignm}, 
$$
\rho \leftarrow \sum_k 
\mathbf{u}_k \mathbf{u}_k^\dagger  \vec \psi (\vec \psi)^\dagger \mathbf{u}_k \mathbf{u}_k^\dagger = 
\sum_k | \mathbf{u}_k^\dagger \vec \psi|^2 \mathbf{u}_k^\dagger \mathbf{u}_k^\dagger  .
$$
This is a mixed state. 
As with ignorant $X$-measurements,  an ignorant measurement in the $Y$-basis alters the quantum state, despite the absence of a specific outcome.

\paragraph{Quantum  HB-POVM}

We get 
$$
\rho \leftarrow \sum_k 
\mathrm{diag}(\mathbf{u}_k)  \vec \psi (\vec \psi)^\dagger \mathrm{diag}(\mathbf{u}^\dagger_k) = 
\sum_k \mathrm{diag}(\mathbf{b}_k \circ \mathbf{p}) =  \mathrm{diag} (\mathbf{p}) .
$$
Thus,  an ignorant measurement in the  $Y$-basis has the same effect as an ignorant measurement in the $X$-basis: We obtain a probabilistic state.

\subsection{Probabilistic Quantum}

Here, we assume that the system is initially in probabilistic state $\mathbf{p}$. 

\paragraph{Probabilistic  PVM}

After performing an ignorant measurement in the $Y$-basis, the posterior state becomes
\begin{equation} \label{eq:pdm2}
\mathbf{p} \;\leftarrow\;
\sum_{k} (\mathbf{b}_k^\top \mathbf{p}) \, \mathbf{b}_k .
\end{equation}
When postselection is applied, the update rule generalizes to
\begin{equation} \label{eq:pdmps2}
\mathbf{p} \;\leftarrow\;
\frac{\sum_{k: z_k = 1} (\mathbf{b}_k^\top \mathbf{p}) \, \mathbf{b}_k}
{\sum_{k: z_k = 1} \mathbf{b}_k^\top \mathbf{p}} .
\end{equation}

\paragraph{Probabilistic  HB-PVM}

Here, using $\mathbf{M}_k^{\mathrm{HB}}$ as defined in Equation~\ref{eq:mhb}, we obtain
\begin{equation} \label{eq:pdmpovmattxxx2}
\mathbf{p} \;\leftarrow\;   \sum_{k}  \mathbf{p} \circ \mathbf{b}_k \;=\; \mathbf{p}.
\end{equation}
Thus, an ignorant measurement leaves the state unchanged—it is effectively ignored.  
With postselection, the update becomes
\begin{equation} \label{eq:pdmpovmattps2}
\mathbf{p} \;\leftarrow\;
\frac{\sum_{k: z_k = 1} \mathbf{p} \circ \mathbf{b}_k}
     {\sum_{k: z_k = 1} \mathbf{p}^\top \mathbf{b}_k} .
\end{equation}

\subsection{Neural  Quantum}
\label{sec:attnqq}

\paragraph{Neural PVM with  Postselection}
For the neural PVM with postselection, the corresponding update of the state vector is given by
\begin{equation} \label{eq:att1pvm} 
\mathbf{q} \;\leftarrow\;
\sum_{k: z_k = 1}
\mathbf{a}_k \,
\mathrm{softmax}^z \!\left(
a_{0,k} + \sum_{\ell=1}^n \gamma_\ell a_{\ell=1,k}
\right).
\end{equation}
The ignorant $Y$-measurement with postselection  can be directly related to the attention mechanism in generative AI.  
Here, the vector $\vec{\gamma}$ serves as the \textit{query}, while the vectors $\mathbf{a}_k$ act simultaneously as \textit{key} and \textit{value}.

\paragraph{Neural HB-POVM with  Postselection}


Starting from Equation~\ref{eq:pdmpovmattps2}  and applying the Hadamard–logistic identity, we obtain
\[
\vec{\gamma} \;\leftarrow\;
\frac{\sum_{k: z_k = 1} \mathbf{p} \circ \mathbf{b}_k}
     {\sum_{k: z_k = 1} \mathbf{p}^\top \mathbf{b}_k}
\;=\;
\frac{
\sum_{k: z_k = 1} \mathbf{p}^\top \mathbf{b}_k \;
\mathrm{sig}(\mathbf{q} + \mathbf{a}_k)
}{
\sum_{k: z_k = 1} \mathbf{p}^\top \mathbf{b}_k
}.
\]
Under Jensen’s approximation, this simplifies to
\[
\vec{\gamma} \;\approx\;
\frac{
\sum_{k: z_k = 1}
\exp\!\left(a_{0,k} + \sum_{\ell=1}^n \gamma_\ell a_{\ell=1,k}\right)
\mathrm{sig}(\mathbf{q} + \mathbf{a}_k)
}{
\sum_{k: z_k = 1}
\exp\!\left(a_{0,k} + \sum_{\ell=1}^n \gamma_\ell a_{\ell=1,k}\right)
}.
\]
Consequently, the neural state update can be written as
\[
\mathbf{q} \;\leftarrow\;
\sum_{k: z_k = 1}
(\mathbf{a}_k + \mathbf{q}) \,
\mathrm{softmax}^z\!\left(
a_{0,k} + \sum_{\ell=1}^n \gamma_\ell a_{\ell=1,k}
\right).
\]
\[
=
\mathbf{q} +
\sum_{k: z_k = 1}
\mathbf{a}_k \,
\mathrm{softmax}^z \!\left(
a_{0,k} + \sum_{\ell=1}^n \gamma_\ell a_{\ell=1,k}
\right).
\]
This is Equation \ref{eq:igg} in Section \ref{sec:igg}.

\section{Appendix: Degenerate and Partial Measurements}
\label{sec:degpam}

This section is relevant for the discussion  on partial measurements in 
Section~\ref{sec:onequbit} and for the discussion of the disjunction fallacy in Section~\ref{sec:intf}. It also shows another way for deriving the attention Equation \ref{eq:att1pvm} in Section \ref{sec:attnqq}.

\subsection{The General Case}

Partial or degenerate measurements reduce the number of possible outcomes from $N$ to $N_C \le N$ by grouping outcomes into clusters.  
We introduce a mapping $c(k) = \tilde k$, indicating that outcome $k$ belongs to cluster $\tilde k$ with $\tilde k \in \{1, \ldots, N_C\}$.  
The effective measurement outcome is thus the cluster index $\tilde k$, and we denote the corresponding random variable by $C$.

In quantum mechanics, this setup corresponds to \textit{degenerate measurements}, in which the observable possesses non-distinct eigenvalues—specifically, all eigenvectors associated with a given cluster share the same eigenvalue.  
By contrast, a \textit{partial measurement} refers to the case in which only a subset of qubits is measured, while the remaining qubits remain in superposition (see Section~\ref{sec:onequbit}).  
Mathematically, however, partial and degenerate measurements are treated identically.  

The probability of obtaining outcome $\tilde k$ is given by
\begin{equation} \label{eq:marg}
P(\tilde k) = \sum_{k: c(k) = \tilde k} |\mathbf{u}_k^\dagger \vec \psi|^2 
\end{equation}
which corresponds to marginalizing over all micro-outcomes $k$ belonging to the same cluster.  
The corresponding pure posterior state is
\begin{equation} \label{eq:aftermeas}
\vec \psi \;\leftarrow\; 
\frac{1}{\sqrt{P(\tilde k)}} 
\sum_{k: c(k) = \tilde k} (\mathbf{u}_k^\dagger \vec \psi) \, \mathbf{u}_k .
\end{equation}
A key property of this update rule is that the prior state $\vec \psi$ influences the posterior through the relative weighting of basis vectors consistent with the degenerate measurement outcome.

When applied to measurements in the $X$-basis, we obtain the partial measurements described in Section~\ref{sec:qubitreprsup}.

\subsection{Disjunction as a Degenerate Measurement}
\label{sec:dijfal}

We now show that the ``OR'' part in the disjunction fallacy
in Section~\ref{sec:intf}
can be implemented as a degenerate measurement.

Assume that $N_C=2$. 
Assume that $c(k')=1$ and $c(k'') = 1$ and for all other outcomes $k$:  $c(k) = 2$.  
If we now have a degenerate measurement with $\tilde k=1$, then (Equation \ref{eq:marg}) 
$$
P(\tilde k=1) =  |\mathbf{u}_{k'}^\dagger \vec \psi|^2  + |\mathbf{u}_{k''}^\dagger \vec \psi|^2 
$$
and (Equation \ref{eq:aftermeas})
\begin{equation} 
\vec \psi \;\leftarrow\; 
\frac{1}{\sqrt{P(\tilde k=1)}} 
 (\mathbf{u}_{k'}^\dagger \vec \psi) \, \mathbf{u}_{k'} 
 +
  (\mathbf{u}_{k''}^\dagger \vec \psi) \, \mathbf{u}_{k''}
 .
\end{equation}
This is a pure state. These results correspond to   Equation~\ref{eq:oreq}.

\subsection{Probabilistic Quantum}
\label{sec:degpc}

We start with probabilistic state $\mathbf{p}$. We have an ignorant measurement in the $Y$-basis, followed by a measurement of cluster variable  $C$. Then, 
\begin{equation} \label{eq:marg-2}
P(\tilde k) = \sum_{k: c(k) = \tilde k} 
\mathbf{b}_k^\top  \mathbf{p} 
\end{equation}

\begin{equation} \label{eq:probaftermeas}
\mathbf{p}\;\leftarrow\; 
\frac{1}{{P(\tilde k)}} 
\sum_{k: c(k) = \tilde k} (\mathbf{b}_k^\top \mathbf{p} ) \, \mathbf{b}_k .
\end{equation}
This is a probabilistic state. 
With only one cluster, $N_C=1$, the last equation becomes identical to an ignorant $Y$-measurement for probabilistic PVM (Equation \ref{eq:pdm2}), which is the basis for the attention Equation \ref{eq:att1pvm}.

\section{Appendix: Quantum Interference in PVM}
\label{sec:intf}

Here we derive the interference equations in Section \ref{sec:qinf}.

Let $P(k') = |\mathbf{u}_{k'}^{\dagger} \vec \psi |^2$, $P(k'') = |\mathbf{u}_{k''}^{\dagger} \vec \psi |^2$. Then,
\begin{equation} \label{eq:oreq}
P(k' \mbox{ OR }  k'') 
= P(k') + P(k'')
\qquad 
\vec \psi \leftarrow  \frac{ (\mathbf{u}_{k'}^{\dagger}  \vec \psi) \mathbf{u}_{k'}
+
 (\mathbf{u}_{k''}^{\dagger}  \vec \psi) \mathbf{u}_{k''} }
 {\sqrt{P(k')  + P(k'')   }}  . 
\end{equation}
We have exploited that outcomes $k'$ and $k''$ are mutually exclusive.  
Then
\[
P(k | k' \mbox{ OR } k'')  = |\mathbf{u}_{k}^{\dagger} \vec \psi|^2
= \frac{| (\mathbf{u}_{k'}^{\dagger}  \vec \psi) (\mathbf{u}_{k}^{\dagger}   \mathbf{u}_{k'})  
+ (\mathbf{u}_{k''}^{\dagger}  \vec \psi )(   \mathbf{u}_{k}^{\dagger}  \mathbf{u}_{k''} ) |^2 }
{P(k')  + P(k'')  }
\]
\[
= \frac{P(k')  P(k|k') +P(k'')  P(k|k'')
+  \mathrm{Interference}
}
{P(k')  + P(k'')  }
\] where  
$P(k|k')  = |\mathbf{u}_k^{\dagger} \mathbf{u}_{k'}   |^2$, $P(k|k'')  = |\mathbf{u}_k^{\dagger} \mathbf{u}_{k''}   |^2$ and
$$\mathrm{Interference} = 2 \mbox{ Re } 
\left(
(\mathbf{u}_{k'}^{\dagger}  \vec \psi) (\mathbf{u}_{k}^{\dagger}   \mathbf{u}_{k'})  
( (\mathbf{u}_{k''}^{\dagger}  \vec \psi )(   \mathbf{u}_{k}^{\dagger}  \mathbf{u}_{k''} ) )^*
\right) .$$ 
We have used that 
$|a+ b|^2 = |a|^2+ |b|^2 + (ab^*) + (ab^*)^* = |a|^2+ |b|^2 + 2 \mbox{ Re }  (ab^*) $.

\section{Appendix: Intractable Neural gHMM}
\label{sec:intghmm}

Here, we derive Equation \ref{eq:bnnu} in Section \ref{sec:intra}. We first consider the case \textit{without} postselection.  
If $\mathbf{B}$ is unitary-stochastic and properly normalized, the outcome probability $P(k)$ follows directly from Equation~\ref{eq:postpure}, and the probabilistic state is updated according to Equation~\ref{sec:posteriorskips}.

\medskip

When postselection is included, we start from the definition in Section~\ref{sec:probposts} and obtain
\[
\bar b_{i(i_1, \ldots, i_n), k} =  
\frac{z_k \exp \!\left( a_{0, k} + \sum_{\ell=1}^n i_\ell a_{\ell=1, k} \right)}
     {\sum_{k: z_k = 1} \exp \!\left( a_{0, k} + \sum_{\ell=1}^n i_\ell a_{\ell=1, k} \right)}
\;\equiv\;
z_k \, \mathrm{softmax}^z \!\left( a_{0, k} + \sum_{\ell=1}^n i_\ell a_{\ell=1, k} \right).
\]
The corresponding outcome probability is then given by
\[
P(k) = 
z_k \sum_{i_1, \ldots, i_n}
\mathrm{softmax}^z \!\left( a_{0, k} + \sum_{\ell=1}^n i_\ell a_{\ell=1, k} \right)
\prod_{\ell=1}^n (\gamma_\ell)^{i_\ell} (1-\gamma_\ell)^{1-i_\ell},
\]
which, under Jensen’s approximation, simplifies to
\[
P(k) \;\approx\;
z_k \, \mathrm{softmax}^z \!\left( a_{0, k} + \sum_{\ell=1}^n \gamma_\ell a_{\ell=1, k} \right).
\]
This is exactly the same expression obtained for the neural HB-POVM with postselection (Equation~\ref{eq:postsk}).  

\medskip

The corresponding posterior distribution takes the form
\begin{equation} \label{eq:bnnuapp}
p_{i(i_1, \ldots, i_n)} \;\leftarrow\;  
\frac{1}{P(k)} \,
\mathrm{softmax}^z \!\left( a_{0, k} + \sum_{\ell=1}^n i_\ell a_{\ell=1, k} \right)
\prod_{\ell=1}^n (\gamma_\ell)^{i_\ell} (1-\gamma_\ell)^{1-i_\ell}.
\end{equation}
Unlike in the neural HB-POVM, where pro-bits remain independent,  
the posterior here introduces dependencies among the pro-bits.  
Consequently, while the neural HB-POVM with postselection remains computationally tractable,  
the full Bayesian update in a generalized HMM (gHMM) with postselection becomes intractable.  
To manage this complexity, variational or related approximate inference methods are typically employed to obtain practical solutions.

\section{Appendix: Experimental Results on Order Effect}
\label{sec:expresul}

\begin{table*}[!h]
\centering
\caption{Comparison of order sensitivity between PVM and HB-POVM on 10k ImageNet samples (200 fine, 16 coarse classes). Divergence metrics (KL, JSD) and label reversal rates reveal that PVM is strongly affected by query order, whereas HB-POVM remains largely invariant.}
\label{tab:pvm_povm}
\renewcommand{\arraystretch}{1.2}
\setlength{\tabcolsep}{5pt}
\begin{tabular}{lcc}
\toprule
\textbf{Metric} & \textbf{PVM} & \textbf{HB-POVM} \\
\midrule
KL & 20.54 & 0.304 \\
JSD & 0.572 & 0.037 \\
Fine Reversal (\%) & 93.56 &  22.86\\
Coarse Reversal (\%) & 30.28 & 15.30 \\
\bottomrule
\end{tabular}
\end{table*}

To evaluate each model’s sensitivity to query order, we query each model twice per input image, once with a \textit{fine-grained} label (e.g., \textit{Golden Retriever, Banana}) and once with a \textit{coarse-grained} label (e.g., \textit{Animal}, \textit{Fruit}). The two possible query orders, \textit{fine$\rightarrow$coarse} and \textit{coarse$\rightarrow$fine}, yield two joint distributions over model predictions. For each image, we record the most probable label (i.e., the class with the maximum predicted probability) from both queries, forming two sets of joint outcomes corresponding to the two query orders. We compute divergence measures between the empirical frequencies of these joint outcomes to assess the model’s consistency under different query orders. The Kullback--Leibler (KL) divergence is $20.54$ for PVM and $0.304$ for HB-POVM, indicating that PVM’s predictions vary substantially across query orders, while HB-POVM remains nearly invariant. A similar trend appears in the Jensen--Shannon divergence (JSD), which symmetrizes the KL measure. We also report \textit{fine} and \textit{coarse reversal rates}, representing the proportion of samples whose fine or coarse labels change between query orders. PVM shows higher reversal rates than HB-POVM, confirming HB-POVM's stronger order invariance.

\begin{table*}[!h]
\centering
\caption{Qualitative example for a dachshund. PVM’s fine and coarse predictions differ across query orders, while HB-POVM preserves consistency.}
\label{tab:roi_comparison}
\renewcommand{\arraystretch}{1.3}
\setlength{\tabcolsep}{8pt}
\begin{tabularx}{\textwidth}{p{3cm} p{3cm} X X}
\toprule
\textbf{ROI} & \textbf{Query Order} & \textbf{PVM Pred.} & \textbf{HB-POVM Pred.} \\
\midrule
\multirow{2}{*}{\includegraphics[width=3cm]{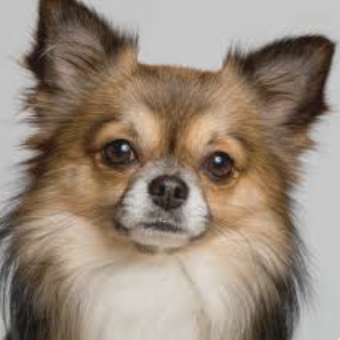}} 
& Fine $\rightarrow$ Coarse & 
Chihuahua, Animal & 
Chihuahua, Animal \\[4.8ex]
\cmidrule(l){2-4}
& Coarse $\rightarrow$ Fine & 
Tarantula, Animal & 
Chihuahua, Animal \\[4.8ex]
\bottomrule
\end{tabularx}
\end{table*}

As a qualitative example, Table~\ref{tab:roi_comparison} presents model predictions for an image of a Chihuahua. Each model is queried twice under different orders, and each cell lists the top-1 fine and coarse predictions. When queried in the \textit{fine$\rightarrow$coarse} order, both models correctly produce \textit{(Chihuahua, Animal)}. However, when the order is reversed, PVM yields an inconsistent pair \textit{(Animal, Tarantula)}. The coarse embedding overrides the internal state and leads the model to predict a fine label aligned with the coarse label rather than the visual input. In contrast, HB-POVM remains stable, producing \textit{(Chihuahua, Animal)} across both query orders.

All experiments are conducted on a subset of ImageNet (\cite{deng2009imagenet}) containing 100,000 samples drawn from 200 fine-grained classes and 16 coarse-grained classes. Each image is annotated with a region of interest (ROI), such as the bounding box of a specific object (e.g., a Chihuahua). The corresponding fine and coarse labels follow the WordNet (\cite{miller1995wordnet}) class hierarchy (e.g., \textit{Chihuahua} $\rightarrow$ \textit{Animal}). ROI visual features are extracted using a ResNet-50 (\cite{he2016deep}) backbone, and the class embeddings of each model are computed as the average of the extracted visual feature vectors.

\end{document}